\documentclass[aps, prl, twocolumn, superscriptaddress, amsmath, 
tightenlines,longbibliography]{revtex4-2}

\usepackage{color}

\usepackage[T1]{fontenc}
\usepackage[english]{babel}

\usepackage{amssymb}
\usepackage{amsmath}
\usepackage{dcolumn}
\usepackage{graphicx}
\usepackage{mathrsfs}
\usepackage{appendix}
\usepackage{graphicx}
\usepackage{booktabs}
\usepackage{colortbl}
\usepackage{verbatim}
\usepackage{pifont}

\definecolor{Dred}{RGB}{190,0,0}

\usepackage{url}
\usepackage[colorlinks]{hyperref}
\hypersetup{%
	plainpages=true,
	breaklinks=true,
	hypertexnames=false, 
	pageanchor=true,
	colorlinks=true,
	linkcolor={blue},
	citecolor={red},
	urlcolor={blue},
	anchorcolor={black}
}

\usepackage{siunitx}

\usepackage{mleftright} %to get rid of the extra space around parentheses with \left and \right: write \mleft, \mright instead

\renewcommand{\eqref}[1]{\mbox{Eq.~(\ref{#1})}}

\newcommand{\figpanel}[2]{Fig.~\hyperref[#1]{\ref*{#1}(#2)}}
\newcommand{\figpanels}[3]{Fig.~\hyperref[#1]{\ref*{#1}(#2)-(#3)}}
\newcommand{\figpanelNoPrefix}[2]{\hyperref[#1]{\ref*{#1}(#2)}}
\newcommand{\ket}[1]{|#1 \rangle}
\newcommand{\bra}[1]{\langle #1|}

\newcommand{\abssq}[1]{\mleft| #1 \mright|^2}

\def \hide#1{}

\begin{document}
	
\title{Nonlinear cascaded quantum network with giant emitters}

\author{Xin Wang}
\email{wangxin.phy@xjtu.edu.cn}
\affiliation{Shaanxi Province Key Laboratory of Quantum Information and Quantum Optoelectronic Devices, School of Physics, Xi'an Jiaotong University, Xi'an 710049, People's Republic of China}
\affiliation{Center for Quantum Computing, RIKEN, Wako-shi, 
Saitama 351-0198, Japan}

\author{Jia-Qi Li}
\affiliation{Shaanxi Province Key Laboratory of Quantum Information and Quantum Optoelectronic Devices, School of Physics, Xi'an Jiaotong University, Xi'an 710049, People's Republic of China}

\author{Zhihai Wang}
\affiliation{Center for Quantum Sciences and School of Physics, Northeast Normal University, Changchun 130024, People's Republic of China}

\author{Anton Frisk Kockum}
\affiliation{Department of Microtechnology and Nanoscience, Chalmers University of Technology, 41296 Gothenburg, Sweden}

\author{Lei Du}
\affiliation{Department of Microtechnology and Nanoscience, Chalmers University of Technology, 41296 Gothenburg, Sweden}

\author{Tao Liu}
\email{liutao0716@scut.edu.cn}
\affiliation{School of Physics and Optoelectronics, South China University of 
Technology, Guangzhou 510640, People's Republic of China}

\author{Franco Nori}
\affiliation{Center for Quantum Computing, RIKEN, Wako-shi, Saitama 351-0198, Japan}
\affiliation{Physics Department, The University of Michigan, Ann Arbor, Michigan 48109-1040, USA}

\date{\today}

%%%%%%%%%%%%%%%%%%%%%%%%%%%%%%%%%%%%%%%%

\begin{abstract}
Chiral quantum optics is central to developing scalable quantum networks, yet existing approaches rely predominantly on linear single-photon regimes. It remains unclear how to generate directional multiphotons. Here we show that giant emitters coupled to nonlinear quantum optical baths enable tunable directional correlated photons, revealing a mechanism for multiphoton directional emission. We demonstrate that the propagation phases of correlated photons, together with the coupling phases of giant emitters, can generate destructive interference in one direction while enhancing emission in the opposite direction, making directionality fully tunable. Building on this mechanism, we introduce a nonlinear cascaded quantum network paradigm mediated by “correlated flying qubits”, providing a configurable building block enabling distinct many-body applications beyond linear unidirectional setups. These results reveal a rich landscape for engineering multiphoton propagation and correlations through interference in giant emitter–nonlinear bath architectures, offering pathways for quantum networks and strongly correlated light–matter platforms.

\end{abstract}

\maketitle

\vspace{.3cm}
\noindent {\large \textbf{Introduction}}
\\
%%%%%%%%%%%%%%%%%%%%%%%%%%%%%%%%%%%%%%%%
Chiral quantum optics, where light propagates in a preferred direction without back-scattering, is a prerequisite for communication protocols in a scalable cascaded quantum network~\cite{Cirac1997, Mitsch2014, Pichler2015, Lodahl2017, Bernardis2023}. Directional photons (serving as flying qubits) can connect remote nodes deterministically~\cite{Guimond2020, Kannan2023}, and underpin unique applications beyond conventional bidirectional networks~\cite{Yao2005,Stannigel2011, Stannigel2012,Vermersch2017, Xiang2017}. Recently, the nonlinear (multiphoton) regime of cascaded quantum systems has yielded many interesting proposals~\cite{Chang2014,Mahmoodian2018, Solano2023} for its potential to provide directional many-body resources in quantum communication, metrology, and sensing, with applications spanning from fundamental science to real-world technology~\cite{Nagata2007, YUAN20101, Giovannetti2011,Monticone2014, Paulisch2019,Sheremet2023}. For example, due to the interplay of nonlinearity and directional transport, many-body ordered states of light can be produced in a waveguide~\cite{Mahmoodian2020}. However, in conventional optical materials, the nonlinearity is ultraweak~\cite{Peyronel2012, Dibyendu2017} and the unidirectional photons are usually uncorrelated. Therefore, constructing nonlinear cascaded quantum networks supporting scalable many-body quantum information processing remains a significant theoretical and experimental challenge, and therefore, has been rarely explored.
 
Recent progress in engineered quantum platforms has opened promising pathways toward overcoming this challenge, such as superconducting circuit QED via the intrinsic anharmonicity~\cite{Koch2007,Orell2019,Blais2021} or Rydberg atom arrays via dipole blockade~\cite{Lukin2001,Gorshkov2011}. A notable example is provided by nonlinear waveguides, which are often modeled by Bose–Hubbard Hamiltonians with on-site photon-photon interactions~\cite{Fedorov2021,Karamlou2024,Claudia2025,Weckesser2025}. In such systems, the interplay between nonlinearity and dispersion leads to strongly correlated multiphoton states, where photons can become bound together in real space~\cite{Winkler2006, Piil2007,Valiente2008,Mansikkamaki2022}. These correlated states, in turn, can mediate supercorrelated radiation dynamics and enable the generation of correlated photon pairs~\cite{WangZhihai2020, Talukdar2022,Li2025,Zhang2025,Rieck2025DoublonBS}.

In parallel, the emerging paradigm of giant atoms has attracted considerable attention~\cite{Anton2014,FriskKockum2020, DuLei2022, Terradas2022, Qiu2023}. A giant atom couples to a photonic or phononic bath at multiple discrete points, with an effective size comparable to the operating wavelength~\cite{Gustafsson2014, Aref2016, Guo2017, Andersson2019, Vadiraj2021, Wang2022, WangZiQi2022}. This multi-point coupling structure gives rise to multiple decay channels that interfere quantum mechanically. By engineering the geometry and phase of these coupling points, one can tailor the interference to facilitate interesting phenomena such as decoherence-free dipole-dipole interactions~\cite{Anton2018, Kannan2020, Carollo2020, Soro2023}, oscillating bound states~\cite{Guo2020, Noachtar2022, Lim2023,Wei2020}, and chiral quantum effects~\cite{Gonzalez2019, Wang2021, Chaitali2023}.

In this work, we show that multiple giant emitters coupled to a nonlinear waveguide enable tunable multiphoton directional emission via interference, revealing a mechanism for the directional emission of strongly correlated photons. Our proposed setup can generate directional correlated photon resources and serve as a building block for constructing nonlinear cascaded quantum networks mediated by ``correlated flying qubits". Furthermore, we demonstrate that this nonlinear network enables numerous many-body applications beyond the capabilities of linear chiral systems.
%----------------------------------------------------------------------------------%
\begin{figure*}
	\centering \includegraphics[width=13cm]{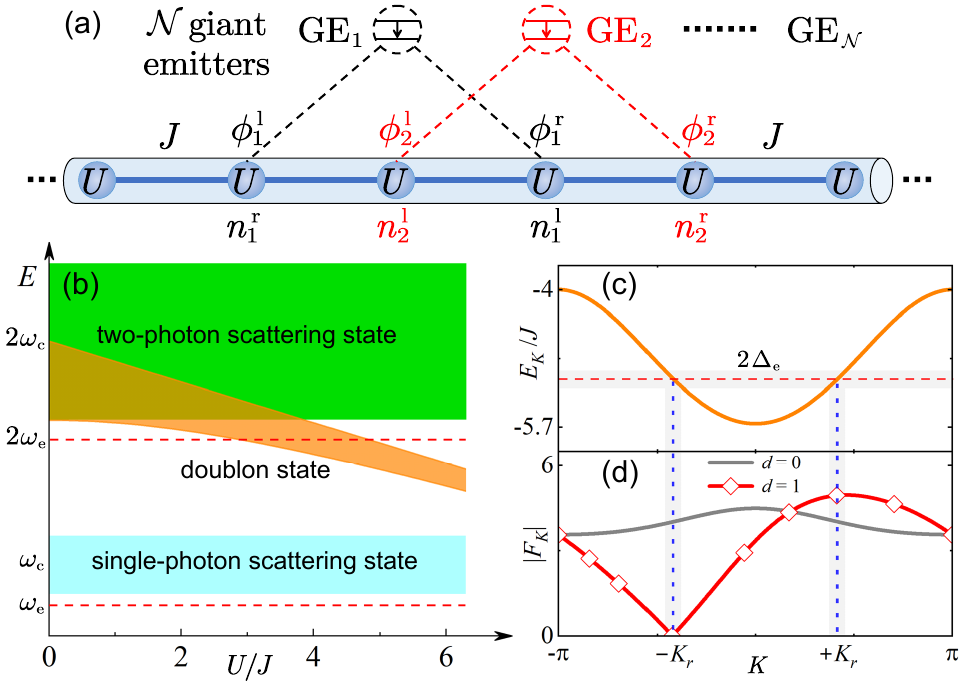}
	\caption{\textbf{Interactions between giant emitters and correlated 
			photons}. (a) Sketch of $\mathcal{N}$ giant emitters (GEs) 
		coupled to a 
		nonlinear waveguide (a coupled-cavity array with hopping rates $J$ 
		and 
		on-site nonlinear potentials $U$). GE $m$ couples to the waveguide at 
		$n_m^{\text{l(r)}}$ with phases $\phi_m^{\text{l(r)}}$. The emitter 
		size is $d = | 
		n_m^{\text{r}} - n_m^{\text{l}} |$; the emitter centers are separated 
		by a distance $D$. 
		(b) Waveguide spectrum as a function of $U$. The lower (upper) red 
		dashed 
		line marks the GE (giant-emitter pair, GEP) frequency 
		$\omega_{\text{e}}$ 
		($2\omega_{\text{e}}$). (c) Doublon dispersion relation $E_K$. The 
		red dashed curve denotes the frequency $2\Delta_{\text{e}}$. (d) 
		Effective 
		coupling strength $|F_K|$ as a function of doublon wave vector $K$ 
		for 
		different $d$. When $d = 1$, $K_r$ is the optimal unidirectional 
		point. Two blue dashed curves denote the resonance mode $\pm K_{r}$.
		Parameters: $D = 0$,  $\Phi^{\text{e}}_{1,2} = \phi_{1,2}^{\text{r}} 
		- \phi_{1,2}^{\text{l}}=\pi / 
		2$, and $U = 4 J$.
		\label{fig1}}
\end{figure*}
%----------------------------------------------------------------------------------%

%%%%%%%%%%%%%%%%%%%%%%%%%%%%%%%%%%%%%%%%

\vspace{.3cm}
\noindent {\large \textbf{Results and Discussion}}
\\
\noindent{\textbf{Model}}
\\
\noindent
The nonlinear waveguide in our setup consists of an array 
of coupled nonlinear cavities, as depicted in Fig.~\ref{fig1}a. In the 
frame rotating with the cavity frequency $\omega_{\text{c}}$, the waveguide 
Hamiltonian is ($\hbar = 1$)
\begin{equation}
H_{\text{w}} = \sum_n \mleft[ \frac{U}{2} a_n^\dag a_n^\dag a_n a_n - J \mleft( 
a_n^\dag a_{n+1} + \text{H.c.} \mright) \mright] ,
\label{Ham_B}
\end{equation}
where $a_n$ ($a^\dag_n$) is the photon annihilation (creation) operator of 
the $n$th cavity, $J$ is the nearest-neighbor hopping rate, and $U$ is the 
on-site photon-photon interaction strength. Throughout this work, we focus on 
systems with attractive interactions $U<0$. A detailed discussion of the 
repulsive case $U>0$ is provided in Supplementary Note 1 and Supplementary 
Note 2 D.

Due to the nonlinear potential,
strongly-correlated photonic states exist in this waveguide. For example, a  
state called ``doublon''~\cite{Winkler2006, Piil2007}, where two 
photons are bound together when propagating along the waveguide, emerges in 
the two-photon subspace. Within the center-of-mass and relative coordinates, 
i.e., $x_{\text{c}} = (n_1 + n_2) / 2$ and $r_{\text{d}} = n_1 - n_2$, the doublon wave function can be 
expressed as $\Psi_K (x_{\text{c}},r_{\text{d}}) = e^{i K x_{\text{c}}} u_K (r_{\text{d}}) $, with $K$ the 
center-of-mass quasi-momentum. The eigen-equation is then derived as 
\begin{gather}
Eu_K(r_{\text{d}})=\!-2J\cos(\frac{K}{2})\sum_{\pm}{u_K(r_{\text{d}}\pm 1)}\!+\!U\delta _{r_{\text{d}},0}u_K(r_{\text{d}}),
\end{gather}
Taking the nonlinear potential as the interaction terms $V(r_{\text{d}})=U\delta_{r_{\text{d}},0}$, the full Green's function is obtained via the Lippmann-Schwinger equation~\cite{Piil2007}:
\begin{gather}
G_K(E,r_{\text{d}}) =G_{K}^{0}(E,\!r_{\text{d}})\!+\!G_{K}^{0}(E,\!r_{\text{d}})\frac{G_{K}^{0}(E,\!r_{\text{d}}=0)U}{1-G_{K}^{0}(E,\!r_{\text{d}}=0)U}. \label{Green_function}
\end{gather}
$G_{K}^{0}(E,r_{\text{d}})$ denotes the unperturbed Green's function, which satisfies $( E - H_0) G_K^0 (E, r_{\text{d}}) = \delta_{r_{\text{d}}, 0}$, with $H_0$ being the unperturbed Hamiltonian without nonlinear terms. Here, we focus on the doublon state, arising from the nonlinear potential $U$. To identify such a bound state, we require
\begin{gather}
1 - G_K^0 (E, r_{\text{d}} = 0) U = 0. \label{dispersion_derive}
\end{gather}
which implies the contribution of the first term in 
Eq.~(\ref{Green_function}), corresponding to scattering states and the 
eigensolution of $H_0$, can be neglected. By solving 
Eq.~(\ref{dispersion_derive}), the dispersion relation of the doublon state 
is derived (see Fig.~\ref{fig1}c) 
\begin{gather}
E_K = - \sqrt{U^2 + 16J^2\cos^2(K/2)}.
\end{gather}
The corresponding relative wave function is given by
\begin{gather}
u_K(r_{\text{d}})=u_0\exp\left[-\frac{|r_{\text{d}}|}{L_u(K)}\right],  \label{relative_postion}
\end{gather}
where $L_{u}^{-1}(K)=\mathrm{asinh}\left[ U/4J\cos(K/2)\right] $ is the inverse 
decay length of the two-photon correlation in space, and $u_0$ is a 
normalization factor. Note that $L_u (K)$ decreases rapidly as the 
interaction strength $U$ increases. This indicates that, in the present of 
the nonlinear potential $U$, the two photons are tightly bound within a 
spatial length $L_{u}$, when propagating along the waveguide. For further 
details on the derivation, we refer the interested reader to 
~\cite{Piil2007} and the Supplementary Note 1.

We consider $\mathcal{N}$ GEs interacting with 
the 
waveguide. 
The two coupling points of GE $m$ are located at $n_m^{\text{l(r)}}$ (see 
Fig.~\ref{fig1}a). The system Hamiltonian is
\begin{equation}
H \!=\! H_{\text{w}} \!+\! \sum_{m=1}^{\mathcal{N}}\! \mleft[ \frac{\Delta_{\text{e}}}{2} \sigma _m^z \!+\! g \!\sum_{\tau 
= \text{l,r}} \mleft( e^{i \phi _m^\tau} \sigma_m^- a_{n_m^\tau}^\dag \!+\! \text{H.c.} 
\mright) \mright],
\label{HTOT}
\end{equation}
with $\Delta_{\text{e}} = \omega_{\text{e}} - \omega_{\text{c}}$ ($\omega_{\text{e}}$ is the emitter frequency), 
$g$ the coupling strength at each coupling point, $\sigma^z$ a Pauli matrix, 
and $\sigma_-$ the lowering operator. 
A local phase $\phi_m^{\text{l(r)}}$ is assumed at each coupling point. 
We first take a giant-emitter pair (GEP), i.e., 
$\mathcal{N}=2$, as an example.
The size of 
emitter $m$ is $d = |n_m^{\text{l}} - n_m^{\text{r}}|$ and the distance between the centers of 
the two emitters is $D = ( |n_2^{\text{l}} + n_2^{\text{r}}| - |n_1^{\text{l}} + n_1^{\text{r}}| ) / 2$. When 
$\Delta_{\text{e}}$ is significantly gapped from scattering states, i.e., $\Delta_{\text{e}} < 
- 2 J$, emission of a single photons is suppressed. We set $2 \Delta_{\text{e}}$ lying 
inside the doublon spectrum (see Fig.~\ref{fig1}b) such that the GEP will 
emit correlated photons, i.e., doublons~\cite{WangZhihai2020}.

%----------------------------------------------------------------------------------%
\begin{figure*}
	\centering \includegraphics[width=\textwidth]{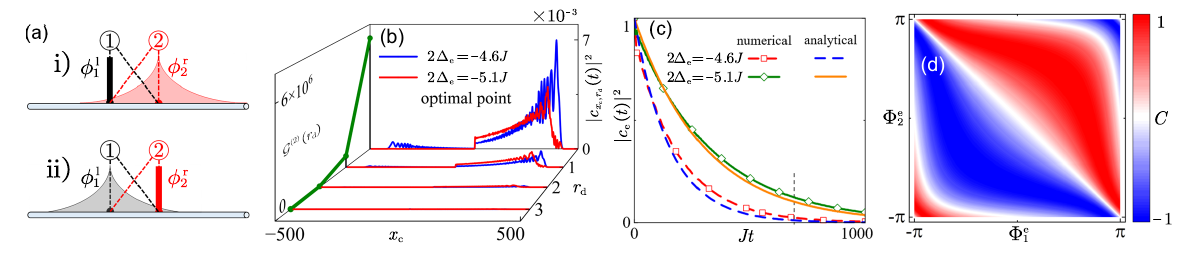}
	\caption{\textbf{Directional emission of doublons}. (a) Sketches of the 
	dual transition processes described by $\vec{e}_{\text{lr}}$. Vertical 
	bars indicate a single photon localized at $n_1^{\text{l}}$ 
	($n_2^{\text{r}}$), while the shaded curves represent the spatial 
	envelope of the intermediate single-photon state. Two giant emitters 
	\{\ding{192}, \ding{193}\} are coupled to the waveguide at positions 
	marked by the black and red dashed arrows, respectively. (b) Two-photon 
	joint probability distribution $|c_{x_{\text{c}},r_{\text{d}}}(t)|^2$ at 
	time $Jt = 600$ (dashed line in c) with $2\Delta_{\text{e}} = -5.1 J$ 
	(optimal emission point). $x_{\text{c}} = (n_1+n_2)/2$ and $r_{\text{d}} 
	= n_1 - n_2$ denote the center-of-mass and relative coordinates of two 
	photons at sites $n_1$ and $n_2$. The two-point correlation function 
	$\mathcal{G}^{(2)}(r_{\text{d}})$ is also shown. (c) Time evolution of 
	$|c_{\text{e}}(t)|^2$ for $2 \Delta_{\text{e}} = - 4.6J$ and $-5.1J$. (d) 
	Chiral factor as a function of $\Phi_{1,2}^{\text{e}} = 
	\phi_{1,2}^{\text{r}}-\phi_{1,2}^{\text{l}}$, with $2\Delta_K = -5.1J$. 
	The coupling strength is set to $g = 0.1 J$; other parameters are the 
	same as in Fig.~\ref{fig1}d.
		\label{fig2}}
\end{figure*}
%----------------------------------------------------------------------------------%

%----------------------------------------------------------------------------------%
\begin{figure}
	\centering \includegraphics[width=8.7cm]{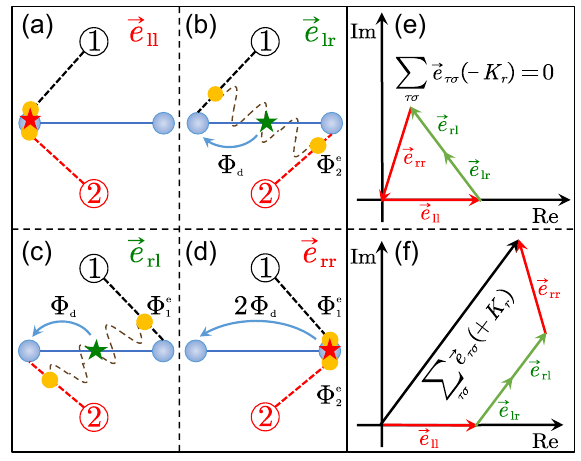}
	\caption{\textbf{Interferences between different decay channels}. (a--d) 
		Sketches of the four decay channels $\vec{e}_{\tau\sigma}$ ($\tau 
		,\sigma 
		= \text{l,r}$) of a GEP releasing two correlated photons into the 
		waveguide. 
		Orange dots represent single photons emitted from individual 
		emitters, 
		while stars denote centers of mass $x_{\text{c}}$ of the correlated 
		photon pairs. 
		(e, f) Summation of $\vec{e}_{\tau \sigma}$ at the optimal point $\pm 
		K_r$ for right- (left-) propagating modes.
		\label{fig3}}
\end{figure}
%----------------------------------------------------------------------------------%

We assume that, initially, the two emitters are in their excited states and 
the waveguide is in its vacuum state. 
Since the Hamiltonian in \eqref{HTOT} 
conserves the total excitation number, the system state is
\begin{align}
	\ket{\psi (t)} =& \mleft[ c_{\text{e}} (t) \sigma_1^+ \sigma _2^+ + \sum_K c_K (t) 
	\mathcal{D}_K^\dag \mright. \notag \\
	&+ \mleft. \sum_{m = 1, 2} \sum_k c_{mk} \sigma_m^+ a_k^\dag \mright] 
	\ket{g, g, \text{vac}} ,
\end{align}
where $c_{\text{e}}(t)$ [$c_K(t)$] is the probability amplitude for the GEP (doublon mode $K$) being excited, $\mathcal{D}_K^\dag$ is the creation operator of doublon mode $K$, $a_k^\dag = 1 / \sqrt{N} \sum_n e^{i k n} a_n^\dag$ are the photonic operators in momentum space, and $|c_{mk}|^2$ is the probability for both the $m$th giant emitter and the single-photon mode $k$ being excited. Given that $g \ll | \Delta_{\text{e}} - 2 J |$, we can adiabatically eliminate the single-photon intermediate states $c_{mk}$ by assuming $\dot{c}_{mk} (t) = 0$. The evolution of $c_{\text{e}}(t)$ and $c_K(t)$ is then given by (see Supplementary Note 2 A for detail discussion)
\begin{align}
i \dot{c}_{\text{e}} (t) &= - \frac{g^2}{J \sqrt{N}} \sum_K F_K^* c_K (t), 
\label{c_e_2} \\
i \dot{c}_K (t) &= \Delta _K c_K (t) - \frac{g^2}{J \sqrt{N}} F_K c_{\text{e}} (t),
\label{c_K_2} 
\end{align}
where $\Delta_K = E_K - 2 \Delta_{\text{e}}$. $F_K$ is expressed as a sum over four decay channels:
\begin{equation}
F_K = \sum_{\tau, \sigma} \vec{e}_{\tau \sigma}(K), \label{FK}
\end{equation}
where each channel is represented as a complex vector to facilitate the interpretation of interference effects. Each vector is defined as
\begin{gather}
\vec{e}_{\tau \sigma}(K) = A_K\left( r_{d}^{\tau \sigma} \right) e^{i \phi_1^\tau} e^{i \phi_2^\sigma} e^{- i K x_{\text{c}}^{\tau \sigma}},
\label{e1}
\end{gather}
with the amplitude 
\begin{gather*}
A_K\left( r_{d}^{\tau \sigma} \right) = \sum_k \cos \mleft[ (k - K/2) r_{d}^{\tau \sigma} \mright] L_{k,K}, \notag \\
L_{k,K} = \frac{2\sqrt{2}J}{N(\omega_k-\Delta_{\text{e}})} \sum_m e^{- i (k - K/2) m} u_K 
(m), \notag 
\end{gather*}
where $x_{\text{c}}^{\tau \sigma} = (n_1^\tau + n_2^\sigma)/2$, $r_{d}^{\tau \sigma} = n_1^\tau - n_2^\sigma$, $\delta_k = \omega_k-\Delta_{\text{e}}$ and $\omega_k = -2J\cos(k)$ is the single-photon dispersion relation. Taking $\vec{e}_{\text{lr}}$ schematically depicted in 
Fig.~\ref{fig2}a as an example, we see that this correlation function 
represents dual processes: emitter 1 (2) excites a single-photon 
intermediate state spreading around the point $n_1^{\text{l}}$ ($n_2^{\text{r}}$) with an 
encoded phase $\phi_1^{\text{l}}$ ($\phi_2^{\text{r}}$). Meanwhile, emitter 2 (1) emits a 
single photon at point $n_2^{\text{r}}$ ($n_1^{\text{l}}$) with encoded phase $\phi_2^{\text{r}}$ 
($\phi_1^{\text{l}}$). The overlap between the wave functions of these photon pairs 
and the doublon state $K$ is proportional to $\vec{e}_{\text{lr}}$ (see Supplementary Note 2 B).

After adiabatically eliminating the single-photon intermediate state, we obtain the reduced two-photon state
\begin{gather}
\ket{\psi (t)} = \mleft[ c_{\text{e}} (t) \sigma_1^+ \sigma _2^+ + \sum_K c_K (t) 
\mathcal{D}_K^\dag \mright] \ket{g, g, \text{vac}}. \label{Reduce_two_photon_space_one_GEP}
\end{gather}
From Eqs.~(\ref{c_e_2}) and~(\ref{c_K_2}), the effective interaction Hamiltonian between GEP and the nonlinear waveguide, i.e., the state $|e,e,\text{vac}\rangle$ and $|g,g,K\rangle$ can be further derived as
\begin{gather}
H_{\mathrm{int}}=\frac{g^2}{J}\sum_{\tau ,\sigma}{\sigma _{1}^{-}}\sigma _{2}^{-}\mathcal{D} _{n_{1}^{\tau},n_{2}^{\sigma}}^{\dagger}+\mathrm{H}.\mathrm{c}., \label{Effective_interaction_Hamiltonian} \\
\mathcal{D} _{n_{1}^{\tau},n_{2}^{\sigma}}^{\dagger}=\frac{1}{\sqrt{N_c}}\sum_K{e^{i\Delta _Kt}}F_K\mathcal{D} _{K}^{\dagger}.
\end{gather}
Note that, the operator $\mathcal{D} _{n_{1}^{\tau},n_{2}^{\sigma}}^{\dagger}$ is defined in the interaction picture, which contains a time-depended phase factor.

%%%%%%%%%%%%%%%%%%%%%%%%%%%%%%%%%%%%%%%%

\vspace{.3cm}
\noindent \textbf{Interference-induced supercorrelated directional 
emission}

\noindent We 
assume that the effective coupling strength $g^2 F_K / J$ is sufficiently 
weak compared to the doublon bandwidth that the Born--Markov approximation 
holds~\cite{Scully1997}. The supercorrelated decay of the GEP then becomes
\begin{equation}
c_{\text{e}} (t) = \exp \mleft( - \frac{\Gamma_+ + \Gamma_-}{2} t \mright), \quad 
\Gamma_{\pm} = \frac{g^4}{v_{\text{g}} J^2} \abssq{F_{\pm K_r}} ,
\label{c_e_Gamma}
\end{equation}
where $E_{K_r} = 2 \Delta_{\text{e}}$, $v_{\text{g}} = \mleft( \partial E_K / \partial K 
\mright) |_{K = K_r}$ is the group velocity, and $\Gamma_+$ ($\Gamma_-$) 
denotes the supercorrelated emission rate to the right (left). Figure~\ref{fig1}d shows $|F_K|$ as a function of $K$ for different $d$. When $d = 0$, the emitters are small and $|F_K| = |F_{-K}|$ always, regardless of $\phi _{1,2}^{\text{r,l}}$, so there is no directional emission. However, for GEs with $d \neq 0$ and $\Phi_{1, 2}^{\text{e}} \neq 0$, directional emission of correlated photon pairs emerges. These rates define the chiral factor $C = (\Gamma _+ - \Gamma _-) / (\Gamma _+ + 
\Gamma_-)$. 

Setting $d = 1$, $D = 0$, and $\Phi^{\text{e}}_{1,2} =  \phi_{1,2}^{\text{r}} - \phi_{1,2}^{\text{l}} = \pi/2$, we plot 
the waveguide 
field distribution and the evolution of $|c_{\text{e}}(t)|^2$ in 
Fig.~\ref{fig2}b, c. The numerically calculated evolution is 
an 
exponential decay, in excellent agreement with analytical 
results. Due to the 
local nonlinear potential, the two photons are strongly bunched; 
the spatial 
correlation function $\mathcal{G}^{(2)}(r_{\text{d}})$ has its maximum at $r_{\text{d}} = 0$. 
Most 
strikingly, the emission of this correlated photon pair exhibits 
a strong 
preference for a certain direction, as shown in 
Fig.~\ref{fig2}b.

Increasing $D$ (fixing $d$), the chiral factor changes little, 
while the 
decay rate quickly decreases to zero. The 
ordering of the 
coupling points has no 
significant effect 
on the directional emission~\cite{Anton2018}. Details on the effect of varying 
$D$ and $d$ are 
given in Supplementary Note 2 C. 
Figure~\ref{fig2}d plots $C$ 
as a function of the phases $\Phi_{1,2}^{\text{e}}$, showing that large 
chiral 
factor is robust to relatively large imprecision in 
$\Phi_{1,2}^{\text{e}}$ and that 
$C$ is smoothly tuned in the full range $C \in (-1, 1)$ by 
changing 
$\Phi_{1,2}^{\text{e}}$. These key results indicate that the proposed 
setup may serve 
well as a source for directional correlated photons. 
%----------------------------------------------------------------------------------%
\begin{figure}
	\centering \includegraphics[width=\linewidth]{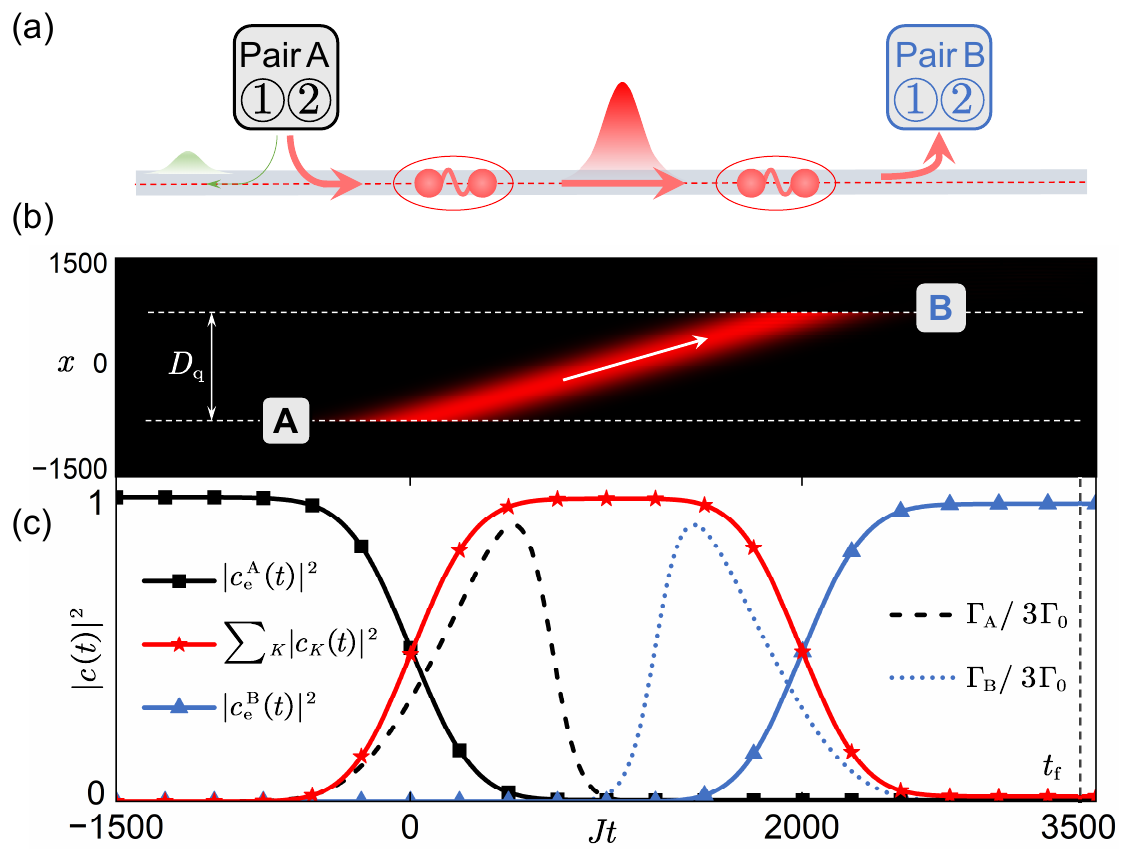}
	\caption{\textbf{State transfer process mediated by directional 
			correlated photons}. (a) Sketch of the nonlinear cascaded quantum 
			system, 
		where correlated photon pairs are unidirectionally emitted by GEP A 
		and 
		subsequently absorbed by GEP B without back-scattering. (b) The 
		evolution 
		of two-photon field distribution at $r_{\text{d}}=0$ and (c) 
		populations 
		for the GEP A (black solid curve with squares), GEP B (blue solid 
		curve 
		with triangles) and doublon (red solid curve with stars) during the 
		two-photon transfer process. The dashed lines in (b) mark the 
		positions 
		of GEP A and B, which are separated by a distance $D_{\text{q}} = 1.5 
		\times 10^3$. The modulations of the couplings are given by 
		Eq.~(\ref{pulse}). The black dashed curve and blue dotted curve 
		denote 
		the modulation pulses of GEP A and B, respectively.	Same parameters 
		as 
		in Fig.~\ref{fig2}.
		\label{fig4}}
\end{figure}
%----------------------------------------------------------------------------------%

To better understand the mechanism for directional emission, we depict the 
four supercorrelated decay channels $\vec{e}_{\tau\sigma}$ in Fig.~\ref{fig3}a--d, with orange 	
dots denoting single photons emitted from distinct emitters and stars being 
the centers-of-mass of correlated photon pairs. The phase of $\vec{e}_{\tau \sigma}$ consists of the local phases $\phi_1^\tau + \phi_2^\sigma$ and the accumulated propagation phase $\Phi_{\text{d}} = - K_r x_{\text{c}}^{\tau \sigma}$ associated with the center-of-mass $x_{\text{c}}^{\tau \sigma}$ of 
the correlated photon pair (not the single-photon propagation phase). Taking $n_1^{\text{l}} = 0$ as the origin, and considering the scenario where two giant emitters couple to the same locations, i.e., $n_1^{\text{l}} = n_2^{\text{l}}$ and $n_1^{\text{r}} = n_2^{\text{r}}$, the four decay channels are simplified as
\begin{align}
\vec{e}_{\text{ll}}= &\ A_{\text{ll}}e^{i\phi _{1}^{\text{l}}}e^{i\phi _{2}^{\text{l}}}=A_{\text{ll}}e^{i\Phi _0}, \notag 
\\
\vec{e}_{\text{lr}}= &\ A_{\text{lr}}e^{i\phi _{1}^{\text{l}}}e^{i\phi _{2}^{\text{r}}}e^{-iK\frac{d}{2}}=A_{\text{lr}}e^{i\Phi _0}e^{i\Phi _{2}^{\text{e}}}e^{-iK\frac{d}{2}}, \notag 
\\
\vec{e}_{\text{rl}}= &\ A_{\text{rl}}e^{i\phi _{1}^{\text{r}}}e^{i\phi _{2}^{\text{l}}}e^{-iK\frac{d}{2}}=A_{\text{rl}}e^{i\Phi _0}e^{i\Phi _{1}^{\text{e}}}e^{-iK\frac{d}{2}}, \notag 
\\
\vec{e}_{\text{rr}}= &\ A_{\text{rr}}e^{i\phi _{1}^{\text{r}}}e^{i\phi _{2}^{\text{r}}}e^{-iKd}=A_{\text{rr}}e^{i\Phi _0}e^{i\Phi _{1}^{\text{e}}}e^{i\Phi _{2}^{\text{e}}}e^{-iKd}, \notag 
\end{align}
where $\Phi_0 = \phi_1^{\text{l}} + \phi_2^{\text{l}}$ is the global reference phase, which 
does not affect the dynamics. The interference is solely determined by 
the phase difference between the two coupling points of each giant emitter, 
i.e., $\Phi_1^{\text{e}} = \phi_1^{\text{r}} - \phi_1^{\text{l}}$ ($\Phi_2^{\text{e}} = \phi_2^{\text{r}} - \phi_2^{\text{l}}$). 
With $n_1^{\text{l}} = n_2^{\text{l}} = 0$ as the reference point, the propagation phases for 
the channels are $\text{ll} \rightarrow \Phi_{\text{d}} = 0$, $\text{lr}, \text{rl} \rightarrow \Phi_{\text{d}} = 
K_r d / 2$, and $\text{rr} \rightarrow \Phi_{\text{d}} = K_r d$. The interferences are 
illustrated in Fig.~\ref{fig3}e, f for $\pm K_r$, respectively. At the 
optimal point $2 \Delta_{\text{e}} = E_{K_r}$, the 
interferences for the left- and right-propagating modes become asymmetric: 
the four vectors for $- K_r$ form a closed loop, i.e.,
$$|\sum_{\tau \sigma} \vec{e}_{\tau \sigma} (+ K_r)| \gg |\sum_{\tau \sigma} 
\vec{e}_{\tau \sigma} (- K_r)| = 0.$$
The interaction between the GEP and the left-propagating mode vanishes (see 
Fig.~\ref{fig1}d) and the GEP only emits correlated photons into the right 
direction, yielding $C \simeq 1$. When $2 \Delta_{\text{e}}$ is biased away from the optimal point, $C$ decreases, as indicated in Fig.~\ref{fig2}d. In the white region where $C=0$, the GEP dynamics recover the small-atom behaviour with non-directional emission~\cite{WangZhihai2020}. The specific parameter regimes corresponding to $C=0$ are discussed in Supplementary Note 2 C. Moreover, the effects of emitter frequency mismatch are discussed in Supplementary Note 2 D.

Note that our examples are the simplest giant emitters with two coupling points. For GEPs with $\mathcal{M}$ coupling points per emitter, there are $\mathcal{M}^2$ supercorrelated decay channels interfering with each other, which is much more complicated than single-photon interference in one giant atom~\cite{Anton2014, Ramos2016, Wang2022}.  When $\mathcal{M} \geq 3$, 
realizing broadband directional emission for doublons is possible by 
considering the optimal methods~\cite{Wang2024}. 
%%%%%%%%%%%%%%%%%%%%%%%%%%%%%%%%%%%%%%%%

\vspace{.3cm}
\noindent \textbf{Nonlinear cascaded quantum network}

\noindent 
Analogously to conventional unidirectional quantum setups 
with uncorrelated 
single photons, 
a cascaded system is naturally formed when considering 
multiple nodes 
unidirectionally 
interacting with the same waveguide. Contrasting linear 
setups, such as 
interacting emitter 
pairs~\cite{Guimond2020,Kannan2023}, multiple-level 
atoms~\cite{Gao2024}, and 
superradiance 
of atom ensembles (time-Dicke 
states)~\cite{WangDawei2015}, our proposal 
utilizes 
supercorrelated photons as ``flying qubits", enhancing 
numerous many-body 
applications. 

Figure~\ref{fig4}a depicts a minimal example setup with two 
separate GEPs, A and B, interacting with a common waveguide. We assume that the two GEPs have the same geometric layout but are separated by a distance $D_{\text{q}}$, with $D_{\text{q}}$ sufficiently larger than $L_u (K_r)$. This ensures that supercorrelated emission between emitters in different pairs, i.e., the process $\bra{\text{vac}}\mathcal{D} _{K}^{\dagger}\sigma _{\mathrm{A},1/2}^{-}\sigma _{\mathrm{B},1/2}^{-}\ket{\text{vac}}$, are suppressed. The operator $\sigma_{\alpha, i}^-$ denotes the lowering operator of emitter $\alpha_i$. Under this condition, we consider the initial state: $$\ket{\psi (t=0)} = \sigma _{\mathrm{A},1}^{+}\sigma _{\mathrm{A},2}^{+}\ket{g, g, \text{vac}},$$ which only GEP A is excited, and states $\sigma _{\mathrm{A},1/2}^{-}\sigma _{\mathrm{B},1/2}^{-}\ket{g, g, \text{vac}}$ can be neglected. Accordingly, the reduced system state analogous to \eqref{Reduce_two_photon_space_one_GEP} becomes:
\begin{equation}
\ket{\psi (t)} = \mleft[ \sum_{\alpha}{c_{\text{e}}^{\alpha}(t)\sigma _{\alpha ,1}^{+}\sigma _{\alpha ,2}^{+}} + \sum_K c_K (t) 
\mathcal{D}_K^\dag \mright] 
\ket{g, g, \text{vac}}.
\label{Reduce_two_photon_space_two_GEP}
\end{equation}
Here, $\alpha = \text{A, B}$ denotes that supercorrelated radiation is confined within individual GEPs. Analogous to \eqref{Effective_interaction_Hamiltonian}, the interaction Hamiltonian can be expanded in the basis of \eqref{Reduce_two_photon_space_two_GEP} as
\begin{equation}
H_{\mathrm{int}}=\frac{g^2}{J}\sum_{\alpha}{\sum_{\tau ,\sigma}{\sigma _{\alpha ,1}^{-}}}\sigma _{\alpha ,2}^{-}\mathcal{D} _{n_{\alpha ,1}^{\tau},n_{\alpha ,2}^{\sigma}}^{\dagger}+\mathrm{H}.\mathrm{c}.
\end{equation}
In contrast to \eqref{Effective_interaction_Hamiltonian}, the labels now include an additional index $\alpha = \text{A, B}$, where $(\alpha, i, \tau)$ denotes the $i$-th emitter in pair $\alpha$ coupled to the waveguide at position $n_{\alpha, i}^\tau$

Consequently, the interaction 
Hamiltonian is simplified as
\begin{equation}
	H_{\rm int} = \sum_{\alpha = \rm{A, B}} \sum_K \frac{g_{\alpha}^2 e^{i 
	\Delta_K t}}{J 
		\sqrt{N}} F_{K, \alpha} \mathcal{D}_k^\dag S_\alpha^- + \text{H.c.} ,
\end{equation}
where $S_\alpha^- = \sigma_{\alpha, 1}^- \sigma_{\alpha, 2}^-$ is the 
joint-quantum-jump operator for GEP $\alpha$ and $F_{K, \alpha}$ is the 
correlation functions of GEP $\alpha$ [see \eqref{FK}]. The
individual supercorrelated decay rates for the two GEPs are 
$\Gamma_{\pm}^{\alpha} = |g_\alpha|^{4}/ (v_{\text{g}} J^2) 
|F_{\alpha, \pm K_r}|^2$ ($\alpha= \text{A, B}$), where $+$ ($-$) 
represents propagation to the right (left). It naturally corresponds to a 
nonlinear master equation with joint-quantum-jump processes. Given that each 
GEP unidirectionally couples to the waveguide with 
$\Gamma_+^{\alpha} \gg \Gamma_-^{\alpha}$, a correlated photon pair (flying 
	qubits) 
is unidirectionally radiated by GEP A, subsequently 
transferred along the waveguide and absorbed by GEP B
without information backflow, which forms a nonlinear cascaded quantum 
network. The numerical example can be seen in Supplementary Note 2 F, and an application (steady four-partite entangled states between remote GEPs) can be seen in Supplementary Note 3 A.

We now demonstrate a high-fidelity process for entangled-state transfer between two remote quantum 
nodes 
by modulating the decay rates of two GEPs. The modulation pulses 
are (see Supplementary Note 3 B)
\begin{equation}
	\Gamma_{+}^{\rm A} (t) = \Gamma_{+}^{\rm B} (\tau_D - t) = \frac{\exp 
	\mleft( - c 
	t^2 \mright)}{\frac{1}{\Gamma_0} - \sqrt{\frac{\pi}{4 c}} \text{erf} 
	\mleft( t \sqrt{c}  \mright)} ,
	\label{pulse}
\end{equation}
where $c = 1.01 \times \Gamma^2_0 \pi / 2$, with $\Gamma_0$ the decay rate of 
GEP A at $t = 0$, and $\tau_{\text{D}} = D_{\text{q}} / v_{\text{g}}$ is the propagation time from A to 
B. The modulation pulses in \eqref{pulse} minimize irreversible quantum jump 
processes and tailor the wavepacket of the correlated photons with 
time-reversal symmetry. Therefore, GEP B will perfectly absorb two correlated 
photons emitted by GEP A~\cite{Cirac1997,Yao2005,Stannigel2011}. Because $\Gamma_{+}^{\alpha} 
(t)
\propto |g_{\alpha}|^4$ in our proposal [cf.~\eqref{c_e_Gamma}], the 
modulated decay rates can be exactly mapped to time-dependent interactions 
$g_{\rm A, \rm B}(t)$. We numerically simulate two-excitation transfer 
between two GEPs in Fig.~\ref{fig4}b,c. Given that the decay 
rates are 
modulated according to Eq.~(20), the dark-state 
conditions hold	in this 
nonlinear cascaded 
system, and the correlated-two-photon field is 	trapped 
between the two GEPs 
without	
leaking outside. At the final time $t_f$, one finds that 
$\{ |c_{\text{e}}^{\rm 
A}(t_f)|^2, \sum_K 
|c_K(t_f)|^2 \} \simeq 0$, and the absorption efficiency 
is $|c_{\text{e}}^{\rm 
B}(t_f)|^2 \simeq 
\SI{98}{\percent}$. This value is primarily limited by the intermediate single-photon process and the finite transfer time~\cite{Stannigel2011}. Reducing the coupling strength during state transfer suppresses the intermediate single-photon amplitudes, thereby improving the corresponding fidelity. However, it also reduces the doublon emission rate, leading to a longer transfer time. This not only slows down the quantum information protocol but also enhances the impact of other decoherence channels. Consequently, in a realistic experimental implementation, an optimal coupling strength $g(t)$ might be chosen to balance transfer fidelity with operation speed.

In linear waveguide QED setups, a single 
multi-level emitter can 
be 
adiabatically reduced to a single effective qubit for 
correlated-photon 
emission~[69]. 
Since such a node lacks internal multipartite structure, 
it cannot support 
the direct transfer of many-body entangled 
	states (e.g., GHZ states) 
	between distant nodes. Moreover, unlike conventional GHZ-state 
distribution in linear 
	waveguides~\cite{Zhong2021}, which 
	requires sequential single-photon transmissions and 
	numerous local 
	operations, our nonlinear cascaded interface 
	accomplishes this many-body 
	task without the need for sequential control or 
	post-processing. For 
	example, an 
initial entangled state of GEP A, $\ket{\Psi_{\rm 
A}(t_0)} = c_{\text{e}}^{\rm A} 
\ket{e, e} + 
c_{\text{g}}^{\rm A} \ket{g, g},$ can be directly mapped onto GEP 
B as $\ket{\Psi_{\rm 
B}(t_0)} = 
c_{\text{e}}^{\rm B} \ket{e, e} + c_{\text{g}}^{\rm B} \ket{g, g}$ in a 
single-step operation.

%----------------------------------------------------------------------------------%
\begin{figure*}
	\centering \includegraphics[width=\textwidth]{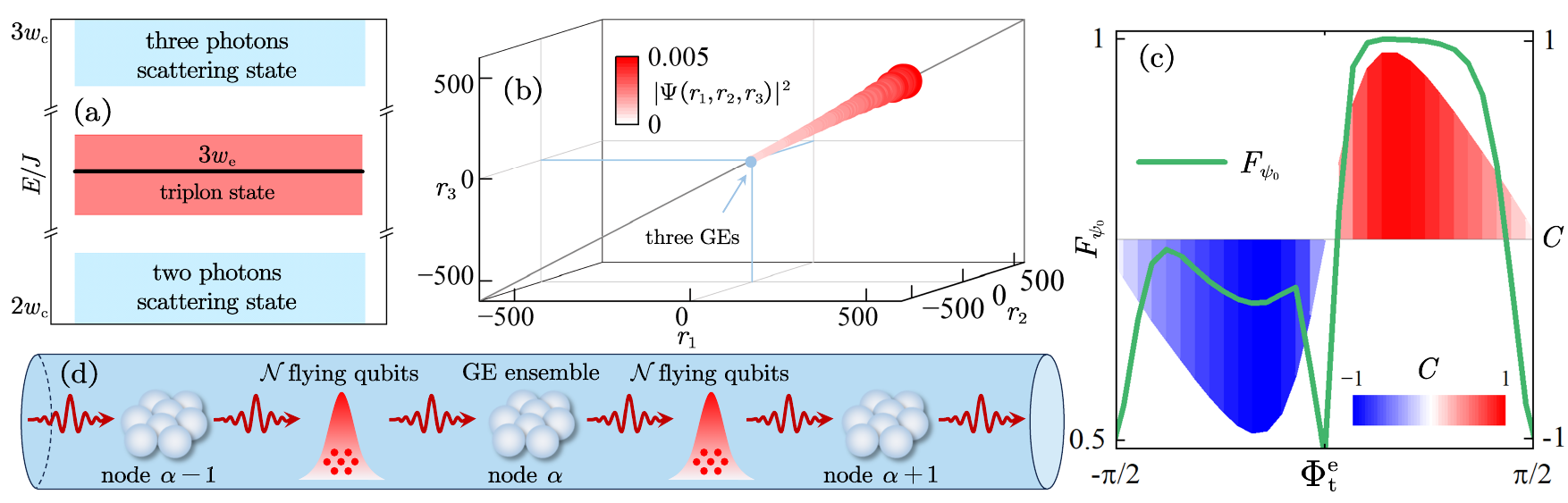}
	\caption{\textbf{Nonlinear cascaded quantum network for many-body 
			applications.} (a) The total frequency of three GEs 
		$3\omega_{\text{e}}$ (the black solid line), is set 
		to be within the triplon spectrum. (b) Field distribution  $|\Psi 
		(r_1,r_2,r_3)|^2$ (mapped as bubble color and size) for triplons 
		emitted 
		from three GEs by setting $\Phi_{\text{t}}^{\text{e}}=0.18\pi$. (c) 
		The Chiral 
		factor $C$ and the trapping fidelity $F_{\psi_0}$ (green solid curve) 
		for the state in 
		Eq.~(\ref{psi0}) as functions of the phase difference 
		$\Phi^{\text{e}}_{\rm t}$. 
		(d) Sketch of a nonlinear cascaded quantum system, whose nodes are 
		unidirectionally connected by $\mathcal{N}$ correlated photons. Here 
		we 
		set $3\Delta_{\text{e}}=-12.7J$ (inside the triplon spectrum). Other 
		parameters 
		are the same as in Fig.~\ref{fig2}.
		\label{fig5}}
\end{figure*}
%----------------------------------------------------------------------------------%

States beyond doublons with $\mathcal{N}$ correlated photons, e.g., 
triplons, also exist in this nonlinear setup~\cite{Strand2015,Sajna2016}.
For example, the energy spectrum of the waveguide in the three-photon
subspace is shown in Fig.~\ref{fig5}a. There, three-photon 
bound states, referred to as ``triplons'', emerge between
two- and three-photon scattering states. 
Those  $\mathcal{N}$ correlated photons can also be unidirectionally routed 
among GEs by exploiting interference. We 
demonstrate this by considering three excited GEs coupling to the waveguide 
with identical phase differences $\Phi^{\text{e}}_{\rm t}$. 
Figure~\ref{fig5}b shows that the three-photon field $|\Psi 
(r_1,r_2,r_3)|^2$ is unidirectionally emitted and strongly bunched at the 
diagonal positions. The joint-quantum-jump operator of this correlated 
process is
$S_\alpha^- = \sigma_{\alpha 1}^- \sigma_{\alpha 2}^-\sigma_{\alpha 3}^-$.
The chiral factor also varies in the full range by changing  $\Phi^{\text{e}}_{\rm t}$ 
(see Fig.~\ref{fig5}c). 

We further consider an application in which the three GEs in two remote nodes are pumped via three-photon parametric down-conversion processes, i.e., 
$H_{\mathrm{pump}} = 
\Omega_{\text{d}}\sum_{\alpha=A,B}(\mathcal{S}_{\alpha}^{-}+\mathcal{S}_{\alpha}^{+}).$
Such a nonlinear pump has been realized, for example, in circuit-QED 
platforms~\cite{Chang2020x}. The dark state for this driven-dissipative 
cascaded system is (see Supplementary Note 4)
\begin{gather}
\ket{\psi_0} = \ket{g}^{\otimes 2\mathcal{N}} + i \frac{2\sqrt{2} 
	\Omega_{\text{d}}}{\Gamma^\alpha_+} \ket{S}, \label{psi0}\\
\ket{S} = \frac{|g\rangle ^{\otimes \mathcal{N}}\otimes |e\rangle 
	^{\otimes \mathcal{N}}-|e\rangle ^{\otimes \mathcal{N}}\otimes 
	|g\rangle 
	^{\otimes \mathcal{N}}}{\sqrt{2}},
\end{gather}
with $\mathcal{N}=3$ for triplons. In the limit $\Omega_{\text{d}} \gg \Gamma^\alpha_+$, the dark state is approximately 
$\ket{\psi_0}\simeq \ket{S}$, which is a six-partite maximally entangled 
state for two remote nodes.  In Fig.~\ref{fig5}c we numerically show that 
the trapping 
fidelity $F_{\psi_0}$ strongly depends on the 
chiral factor. In an ideal nonlinear cascaded network with $C \simeq 1$, the 
trapping fidelity can reach as high as $F_{\psi_0} \simeq 1$. When 
this nonlinear system transitions to a bidirectional configuration, 
$F_{\psi_0}$
decreases accordingly. 

Furthermore, we infer that groups of $\mathcal{N}$-photon states in nonlinear waveguides can serve as directional ``correlated flying qubits''. Figure~\ref{fig5}d schematically depicts a cascaded 
quantum network where $\mathcal{N}$ GEs serve as quantum nodes, whose 
joint-quantum-jump operators are
$\mathcal{S} _{\alpha}^{-}=\prod_j^{\mathcal{N}}{\sigma_{\alpha j}^{-}}$.
Analogously, the cascaded master equation is 
\begin{gather}
\dot{\rho}=-i\left(H_{\mathrm{eff}}\rho-H_{\mathrm{eff}}^{\dagger}\rho 
\right) +\mathcal{L} \rho \mathcal{L} ^{\dagger}, \\
H_{\mathrm{eff}}=-i 
\sum_{\alpha}{\frac{\Gamma^\alpha _+}{2}\mathcal{S}_{\alpha}^{+}\mathcal{S} 
_{\alpha}^{-}}-i\sum_{\alpha \beta}^{\beta >\alpha}\sqrt{\Gamma^\alpha 
_+\Gamma^\beta _+}{\mathcal{S} 
_{\beta}^{+}\mathcal{S} _{\alpha}^{-}} ,
\end{gather}
where
$\mathcal{L} =\sum_{\alpha}{\sqrt{\Gamma^\alpha _+}\mathcal{S} 
_{\alpha}^{-}}$. Following the methods in Fig.~\ref{fig4}, our proposal enable single-step transfer of an arbitrary $\mathcal{N}$-qubit GHZ state, $|\psi _0\rangle = c_{\text{e}}|e\rangle ^{\otimes \mathcal{N}}+c_{\text{g}}|g\rangle ^{\otimes \mathcal{N}}$, between two $\mathcal{N}$-body nodes, without decomposing it into sequential single-excitation processes. In summary, our work demonstrates that nonlinear waveguides can be configured as multiphoton buses supporting advanced many-body quantum applications in future study.

\vspace{.3cm}
\noindent \textbf{Feasible implementation}

\noindent 
Our proposal is feasible in circuit-QED systems~\cite{Gu2017, Krantz2019, Blais2021}. An array of coupled transmon qubits~\cite{Koch2007} can be configured as the nonlinear waveguide. The hopping rate and Kerr nonlinearity of such an array demonstrated in experiments~\cite{Gourgy2015, Roushan2017, Fedorov2021, Kim2021} are on the order of $J / 2 \pi \sim \SI{50}{\mega\hertz}$ and $U / 2 \pi \sim \SI{200}{\mega\hertz}$, respectively. 

The giant-atom configuration, with multiple discrete coupling points to a waveguide, has already been experimentally realized~\cite{Kannan2020, Vadiraj2021}. In our scheme, the relative coupling phases required for directional emission can be dynamically tuned via parametric modulation~\cite{Chaitali2023}. The transmon anharmonicity $U_{\text{giant}} / 2 \pi \sim \SI{300}{\mega\hertz}$ supports the two-level approximation for the emitters. Furthermore, time-dependent interactions between superconducting giant atoms and the nonlinear waveguide can be realized using the methods~\cite{Roushan2017nature,Chaitali2023}, achieving a coupling strength of $g/2\pi=0.1J/2\pi \sim \SI{5}{\mega\hertz}$. 

Under these parameters, the supercorrelated emission rate is estimated as $\Gamma_{\pm} / 2 \pi \sim \SI{1}{\mega\hertz}$ (see Supplementary Note 3 A), which is much larger than the intrinsic decoherence rate of a circuit-QED platform~\cite{Krinner2022m, Place2021m}. Therefore, the phenomena predicted in this work are possible to observe in current circuit-QED setups.

%%%%%%%%%%%%%%%%%%%%%%%%%%%%%%%%%%%%%%%%
\vspace{.3cm}
\noindent {\large\textbf{Conclusions}}
\\
\noindent
We show how to realize unidirectional transport of strongly 
correlated photons in a nonlinear waveguide interacting with GEs. Compared to 
the single-photon case, the interference channels are more complicated. The 
propagating phase related to the center of mass correlated photons plays an 
crucial role in interference between those channels. Another key ingredient, 
the local coupling phases of GEs, does not only lead to asymmetric 
interference relations for oppositely propagating modes, but also provides 
the necessary degree of freedom for tuning the emission directions. 
Our setup can be a building block for generalized cascaded 
quantum networks with ``correlated flying qubits'', enabling numerous 
applications which are not accounted for in a linear cascaded interface.

Our work demonstrates an example of how correlated emission is significantly modified by the interplay of giant atoms and the bath nonlinearity. We believe it offers perspectives for harnessing many-body states, and thus provides a powerful toolbox for quantum-enhanced metrology, quantum information processing, and simulations~\cite{Nagata2007,YUAN20101, Giovannetti2011, Monticone2014, Paulisch2019}.

%%%%%%%%%%%%%%%%%%%%%%%%%%%%%%%%%%%%%%%%
\vspace{.3cm}
\noindent {\large \textbf{Methods}}
\\
\noindent \textbf{Numerical simulation}
\\
\noindent
To check the validity of the theoretical results in this study, we perform numerical simulations by considering an open-boundary waveguide with length $N = 3000$ in real space. In the two-excitation subspace, the dimension of this Hilbert space is approximately $N_{\text{H}} \simeq N^2/2 = 5 \times 10^6$. The numerical simulations of quantum dynamics are based on the open-source Python package QuTiP ~\cite{Johansson12qutip, Johansson13qutip}, which
agree well with theoretical results throughout our work.

\vspace{.5cm}
\noindent {\textbf{Nonlinear cascaded master equation}}
\\
\noindent
We demonstrate how to derive the nonlinear cascaded master equation using the example of doublons serving as "flying qubits." Specifically, the evolution governed by the interaction Hamiltonian in Eq. (18) is derived as follows:
\begin{align}
\dot{\rho} = &- \frac{i}{\hbar} {\rm{Tr}_{\rm R}} \mleft[ H_{\rm int} (t), \rho (t) \mright] 
\notag \\
&- \frac{1}{\hbar^2} {\rm{Tr}_{\rm R}}\! \int_{t_0}^t\!\!dt'\ \mleft[ H_{\rm int} (t), \mleft[ H_{\rm int} (t'), \rho_S (t') \otimes \rho_R \mright] \mright] ,
\end{align}
where $\rho_S$ ($\rho_R$) denotes the density matrix of the emitters (nonlinear waveguide).
The correlation relations for a doublon in the waveguide satisfy $\langle \mathcal{D} _K\rangle =\langle \mathcal{D} _{K}^{\dagger}\rangle =\langle \mathcal{D} _{K}^{\dagger}\mathcal{D} _K\rangle =0
$ and $\langle \mathcal{D} _K\mathcal{D} _{K}^{\dagger}\rangle =1$. Then, we 
obtain
\begin{equation}
\dot{\rho} \!=\! -\!\!\! \int_{t_0}^t\!\! dt'\!\sum_{\alpha, \beta}\! \mathscr{A}_{\alpha, \beta} \! \mleft[ - \! S_\beta^-  \rho_S \left(t'\right) S_\alpha^+ \!+\! S_\alpha^+S_\beta^-  \rho_S (t') \mright] \!+\! \text{H.c.} ,
\label{rho_1}
\end{equation}
where the coefficient is
\begin{equation}
\mathscr{A}_{\alpha, \beta} = \frac{g^{*2}_\alpha g^{2}_\beta}{J^2} 
\frac{1}{N_c} \sum_K 
F_{K, \alpha}^* F_{K, \beta} e^{- i \Delta_K (t - t')} .
\end{equation}
Since only the modes centered around $K_r$ are excited with high probability, the summation over $K$ can be approximated as
\begin{align}
&\sum_K F_{K, \alpha}^* F_{K, \beta} e^{- i \Delta_K (t - t')} 
\notag \\
&\simeq \frac{N_c}{2 \pi} F_{K_r, \alpha}^* F_{K_r, \beta} \int_{-\pi}^\pi dK\ e^{- i (K - K_r) v_{\text{g}} \mleft( t - t' - \frac{n_\alpha - n_\beta}{v_{\text{g}}} \mright)} 
\notag \\
&= \frac{N_c}{v_{\text{g}}} F_{K_r, \beta}^* F_{K_r, \alpha} \delta \mleft( t - t' - \frac{\Delta x_{\text{c}}}{v_{\text{g}}} \mright) ,
\end{align}
where $\Delta x_{\text{c}} / v_{\text{g}} = (n_\alpha - n_\beta) / v_{\text{g}}$ is the time delay between the two pairs, corresponding to the propagation time between pairs A and B with $n_{\alpha / \beta} = \mleft( n_{\alpha / \beta, 1}^{\text{r}} - n_{\alpha / \beta, 1}^{\text{l}} \mright) / 2$ being the positions of GEPs $\alpha / \beta$. By employing the properties of the $\delta$ function,
\begin{equation}
\int_{t_0}^t \!dt' \delta \mleft(\! t \!-\! t' \!-\! \frac{\Delta x_{\text{c}}}{v_{\text{g}}} \! \mright)  \rho_S \!\left(t'\right)\! = \!\!  
\begin{cases}
\!0 \qquad \quad \,\, \Delta x_{\text{c}} / v_{\text{g}} \!<\! 0, \\
\!\frac{1}{2} \rho_S (t) \quad \Delta x_{\text{c}} / v_{\text{g}} \!=\! 0, \\
\!\rho_S (t) \quad \,\,\,\, \Delta x_{\text{c}} / v_{\text{g}} \!>\! 0, 
\end{cases}
\label{deltafunction}
\end{equation}
the coefficients are written as ($n_{\rm A} < n_{\rm B}$)
\begin{gather}
\mathscr{A} _{\alpha 
,\alpha}\!=\!\frac{1}{2}\frac{g_{\alpha}^{4}}{J^2v_{\text{g}}}\!\sum_{\pm}{\!}|F_{\pm 
K_r,\alpha}|^2\!=\!\frac{\Gamma _{+}^{\alpha}\!+\Gamma 
_{-}^{\alpha}}{2}, \!\! \!\quad \alpha\!=\!\rm{A}\!/\ \!\!\rm{B}, 
\\	
\mathscr{A}_{\rm{A}, \rm{B}} \!=\! \frac{g^{*2}_\alpha g^{2}_\beta}{J^2v_{\text{g}}} 
F_{-K_r, \rm{A}}^* 
F_{-K_r, \rm{B}} \!=\! \Gamma _{-}^{\mathrm{AB}}
, \\
\mathscr{A}_{\rm{B}, \rm{A}} \!=\! \frac{g^{2}_\alpha g^{*2}_\beta}{J^2v_{\text{g}}} 
F_{+K_r, \rm{B}}^* 
F_{+K_r, \rm{A}} \!=\! \Gamma _{+}^{\mathrm{BA}},
\end{gather}
where we neglect the time-retardation effects by assuming $\rho (t - \Delta 
x_{\text{c}} / v_{\text{g}}) \approx \rho (t)$, and $\Gamma^{\alpha}_{\pm}$ correspond to 
the decay 
rates into the right- and left-propagating modes, respectively. 

Finally, the master equation is
\begin{align}
\dot{\rho} =& - \sum_{\alpha = \rm{A} / \rm{B}}
\frac{\Gamma _{+}^{\alpha }+\Gamma _{-}^{\alpha}}{2}
\mleft[ - S_\alpha^- \rho_S (t) S_\alpha^+ + S_\alpha^- S_\alpha^+ \rho_S (t) 
\mright] \notag \\
&- \Gamma _{+}^{\mathrm{BA}} \mleft[ - S_{\rm A}^- \rho_S (t) S_{\rm B}^+ + 
S_{\rm B}^+ S_{\rm A}^- \rho_S (t) \mright] \notag \\
&- \Gamma _{-}^{\mathrm{AB}} \mleft[ - S_{\rm B}^- \rho_S (t) S_{\rm A}^+ + 
S_{\rm A}^+ S_{\rm B}^- \rho_S (t) \mright] + \text{H.c.}
\end{align}
Note that $S_{\rm A}^- = \sigma_{\rm{A}, 1}^- \sigma_{\rm{A}, 2}^-$ ($S_{\rm 
B}^- = \sigma_{\rm{B}, 1}^- \sigma_{\rm{B}, 2}^-$) are joint-quantum-jump 
operators describing the two emitters in GEP A (B) simultaneously jumping to 
their ground states. 
Given that the two GEPs only interact with 
right-propagating modes, i.e., $\Gamma^{\alpha}_+ \gg 
\Gamma^{\alpha}_- \simeq 0$, the master equation can be written as 
\begin{align}
\dot{\rho}&=-i\left( H_{\mathrm{eff}}\rho -\rho H_{\mathrm{eff}}^{\dagger} 
\right) +\mathcal{L} \rho \mathcal{L} ^{\dagger}, \label{cascaded3}\\
H_{\mathrm{eff}}&=-i\sum_{\alpha =\mathrm{A},\mathrm{B}}{\frac{\Gamma 
_{+}^{\alpha}}{2}S_{\alpha}^{+}S_{\alpha}^{-}}-i\sqrt{\Gamma 
_{+}^{\mathrm{A}}\Gamma 
_{+}^{\mathrm{B}}}S_{\mathrm{B}}^{+}S_{\mathrm{A}}^{-}, \\
\mathcal{L} &= \sqrt{\Gamma^{\mathrm{A}}_+}S_{\rm A}^- + 
\sqrt{\Gamma^{\mathrm{B}}_+}S_{\rm B}^- .
\label{cascaded1}
\end{align}
where we have used the relation $\Gamma _{+}^{\mathrm{BA}}=\sqrt{\Gamma 
_{+}^{\mathrm{A}}\Gamma _{+}^{\mathrm{B}}}$.
The second term of $H_{\rm eff}$ is unique to the cascaded system. It 
describes the process where a correlated photon pair is initially radiated by 
GEP A and subsequently absorbed by the other emitter pair without 
back-scattering.

%%%%%%%%%%%%%%%%%%%%%%%%%%%%%%%%%%%%%%%%
\vspace{.3cm}
\noindent {\large \textbf{Data availability}}
\\
\noindent
Supplementary data for all the figures are provided at 
~\url{https://doi.org/10.5281/zenodo.18409019}. Additional data that support 
the findings of this study are available from the authors upon request.

\vspace{.3cm}
\noindent {\large \textbf{Code availability}}
\\
\noindent
The codes used for the simulation and analysis of the data are available from 
the authors upon request.

\vspace{.3cm}
\noindent {\large \textbf{References}}
\\
\noindent
%%%%%%%%%%%%%%%%%%%%%%%%%%%%%%%%%%%%%%%%

%\bibliography{doublon_chiral_ref}

\begin{thebibliography}{84}%
	\makeatletter
	\providecommand \@ifxundefined [1]{%
		\@ifx{#1\undefined}
	}%
	\providecommand \@ifnum [1]{%
		\ifnum #1\expandafter \@firstoftwo
		\else \expandafter \@secondoftwo
		\fi
	}%
	\providecommand \@ifx [1]{%
		\ifx #1\expandafter \@firstoftwo
		\else \expandafter \@secondoftwo
		\fi
	}%
	\providecommand \natexlab [1]{#1}%
	\providecommand \enquote  [1]{``#1''}%
	\providecommand \bibnamefont  [1]{#1}%
	\providecommand \bibfnamefont [1]{#1}%
	\providecommand \citenamefont [1]{#1}%
	\providecommand \href@noop [0]{\@secondoftwo}%
	\providecommand \href [0]{\begingroup \@sanitize@url \@href}%
	\providecommand \@href[1]{\@@startlink{#1}\@@href}%
	\providecommand \@@href[1]{\endgroup#1\@@endlink}%
	\providecommand \@sanitize@url [0]{\catcode `\\12\catcode `\$12\catcode
		`\&12\catcode `\#12\catcode `\^12\catcode `\_12\catcode `\%12\relax}%
	\providecommand \@@startlink[1]{}%
	\providecommand \@@endlink[0]{}%
	\providecommand \url  [0]{\begingroup\@sanitize@url \@url }%
	\providecommand \@url [1]{\endgroup\@href {#1}{\urlprefix }}%
	\providecommand \urlprefix  [0]{URL }%
	\providecommand \Eprint [0]{\href }%
	\providecommand \doibase [0]{https://doi.org/}%
	\providecommand \selectlanguage [0]{\@gobble}%
	\providecommand \bibinfo  [0]{\@secondoftwo}%
	\providecommand \bibfield  [0]{\@secondoftwo}%
	\providecommand \translation [1]{[#1]}%
	\providecommand \BibitemOpen [0]{}%
	\providecommand \bibitemStop [0]{}%
	\providecommand \bibitemNoStop [0]{.\EOS\space}%
	\providecommand \EOS [0]{\spacefactor3000\relax}%
	\providecommand \BibitemShut  [1]{\csname bibitem#1\endcsname}%
	\let\auto@bib@innerbib\@empty
	%</preamble>
	\bibitem [{\citenamefont {Cirac}\ \emph {et~al.}(1997)\citenamefont 
	{Cirac},
		\citenamefont {Zoller}, \citenamefont {Kimble},\ and\ \citenamefont
		{Mabuchi}}]{Cirac1997}%
	\BibitemOpen
	\bibfield  {author} {\bibinfo {author} {\bibfnamefont {J.~I.}\ 
	\bibnamefont
			{Cirac}}, \bibinfo {author} {\bibfnamefont {P.}~\bibnamefont 
			{Zoller}},
		\bibinfo {author} {\bibfnamefont {H.~J.}\ \bibnamefont {Kimble}},\ 
		and\
		\bibinfo {author} {\bibfnamefont {H.}~\bibnamefont {Mabuchi}},\ 
		}\bibfield
	{title} {\bibinfo {title} {{Quantum State Transfer and Entanglement
				Distribution among Distant Nodes in a Quantum Network}},\ 
				}\href
	{https://doi.org/10.1103/PhysRevLett.78.3221} {\bibfield  {journal} 
	{\bibinfo
			{journal} {Phys. Rev. Lett.}\ }\textbf {\bibinfo {volume} {78}},\ 
			\bibinfo
		{pages} {3221} (\bibinfo {year} {1997})}\BibitemShut {NoStop}%
	\bibitem [{\citenamefont {Mitsch}\ \emph {et~al.}(2014)\citenamefont 
	{Mitsch},
		\citenamefont {Sayrin}, \citenamefont {Albrecht}, \citenamefont
		{Schneeweiss},\ and\ \citenamefont {Rauschenbeutel}}]{Mitsch2014}%
	\BibitemOpen
	\bibfield  {author} {\bibinfo {author} {\bibfnamefont {R.}~\bibnamefont
			{Mitsch}}, \bibinfo {author} {\bibfnamefont {C.}~\bibnamefont 
			{Sayrin}},
		\bibinfo {author} {\bibfnamefont {B.}~\bibnamefont {Albrecht}}, 
		\bibinfo
		{author} {\bibfnamefont {P.}~\bibnamefont {Schneeweiss}},\ and\ 
		\bibinfo
		{author} {\bibfnamefont {A.}~\bibnamefont {Rauschenbeutel}},\ 
		}\bibfield
	{title} {\bibinfo {title} {Quantum state-controlled directional 
	spontaneous
			emission of photons into a nanophotonic waveguide},\ }\href
	{https://doi.org/10.1038/ncomms6713} {\bibfield  {journal} {\bibinfo
			{journal} {Nature Communications}\ }\textbf {\bibinfo {volume} 
			{5}},\
		\bibinfo {pages} {5713} (\bibinfo {year} {2014})}\BibitemShut 
		{NoStop}%
	\bibitem [{\citenamefont {Pichler}\ \emph {et~al.}(2015)\citenamefont
		{Pichler}, \citenamefont {Ramos}, \citenamefont {Daley},\ and\ 
		\citenamefont
		{Zoller}}]{Pichler2015}%
	\BibitemOpen
	\bibfield  {author} {\bibinfo {author} {\bibfnamefont {H.}~\bibnamefont
			{Pichler}}, \bibinfo {author} {\bibfnamefont {T.}~\bibnamefont 
			{Ramos}},
		\bibinfo {author} {\bibfnamefont {A.~J.}\ \bibnamefont {Daley}},\ and\
		\bibinfo {author} {\bibfnamefont {P.}~\bibnamefont {Zoller}},\ 
		}\bibfield
	{title} {\bibinfo {title} {Quantum optics of chiral spin networks},\ 
	}\href
	{https://doi.org/10.1103/PhysRevA.91.042116} {\bibfield  {journal} 
	{\bibinfo
			{journal} {Phys. Rev. A}\ }\textbf {\bibinfo {volume} {91}},\ 
			\bibinfo
		{pages} {042116} (\bibinfo {year} {2015})}\BibitemShut {NoStop}%
	\bibitem [{\citenamefont {Lodahl}\ \emph {et~al.}(2017)\citenamefont 
	{Lodahl},
		\citenamefont {Mahmoodian}, \citenamefont {Stobbe}, \citenamefont
		{Rauschenbeutel}, \citenamefont {Schneeweiss}, \citenamefont {Volz},
		\citenamefont {Pichler},\ and\ \citenamefont {Zoller}}]{Lodahl2017}%
	\BibitemOpen
	\bibfield  {author} {\bibinfo {author} {\bibfnamefont {P.}~\bibnamefont
			{Lodahl}}, \bibinfo {author} {\bibfnamefont {S.}~\bibnamefont 
			{Mahmoodian}},
		\bibinfo {author} {\bibfnamefont {S.}~\bibnamefont {Stobbe}}, \bibinfo
		{author} {\bibfnamefont {A.}~\bibnamefont {Rauschenbeutel}}, \bibinfo
		{author} {\bibfnamefont {P.}~\bibnamefont {Schneeweiss}}, \bibinfo 
		{author}
		{\bibfnamefont {J.}~\bibnamefont {Volz}}, \bibinfo {author} 
		{\bibfnamefont
			{H.}~\bibnamefont {Pichler}},\ and\ \bibinfo {author} 
			{\bibfnamefont
			{P.}~\bibnamefont {Zoller}},\ }\bibfield  {title} {\bibinfo 
			{title} {Chiral
			quantum optics},\ }\href {https://doi.org/10.1038/nature21037} 
			{\bibfield
		{journal} {\bibinfo  {journal} {Nature}\ }\textbf {\bibinfo {volume} 
		{541}},\
		\bibinfo {pages} {473} (\bibinfo {year} {2017})}\BibitemShut {NoStop}%
	\bibitem [{\citenamefont {De~Bernardis}\ \emph {et~al.}(2023)\citenamefont
		{De~Bernardis}, \citenamefont {Piccioli}, \citenamefont {Rabl},\ and\
		\citenamefont {Carusotto}}]{Bernardis2023}%
	\BibitemOpen
	\bibfield  {author} {\bibinfo {author} {\bibfnamefont {D.}~\bibnamefont
			{De~Bernardis}}, \bibinfo {author} {\bibfnamefont {F.~S.}\ 
			\bibnamefont
			{Piccioli}}, \bibinfo {author} {\bibfnamefont {P.}~\bibnamefont 
			{Rabl}},\
		and\ \bibinfo {author} {\bibfnamefont {I.}~\bibnamefont {Carusotto}},\
	}\bibfield  {title} {\bibinfo {title} {{Chiral Quantum Optics in the Bulk 
	of
				Photonic Quantum Hall Systems}},\ }\href
	{https://doi.org/10.1103/PRXQuantum.4.030306} {\bibfield  {journal} 
	{\bibinfo
			{journal} {PRX Quantum}\ }\textbf {\bibinfo {volume} {4}},\ 
			\bibinfo {pages}
		{030306} (\bibinfo {year} {2023})}\BibitemShut {NoStop}%
	\bibitem [{\citenamefont {Guimond}\ \emph {et~al.}(2020)\citenamefont
		{Guimond}, \citenamefont {Vermersch}, \citenamefont {Juan}, 
		\citenamefont
		{Sharafiev}, \citenamefont {Kirchmair},\ and\ \citenamefont
		{Zoller}}]{Guimond2020}%
	\BibitemOpen
	\bibfield  {author} {\bibinfo {author} {\bibfnamefont {P.-O.}\ 
	\bibnamefont
			{Guimond}}, \bibinfo {author} {\bibfnamefont {B.}~\bibnamefont 
			{Vermersch}},
		\bibinfo {author} {\bibfnamefont {M.~L.}\ \bibnamefont {Juan}}, 
		\bibinfo
		{author} {\bibfnamefont {A.}~\bibnamefont {Sharafiev}}, \bibinfo 
		{author}
		{\bibfnamefont {G.}~\bibnamefont {Kirchmair}},\ and\ \bibinfo {author}
		{\bibfnamefont {P.}~\bibnamefont {Zoller}},\ }\bibfield  {title} 
		{\bibinfo
		{title} {A unidirectional on-chip photonic interface for 
		superconducting
			circuits},\ }\href {https://doi.org/10.1038/s41534-020-0261-9} 
			{\bibfield
		{journal} {\bibinfo  {journal} {npj Quantum Information}\ }\textbf 
		{\bibinfo
			{volume} {6}},\ \bibinfo {pages} {32} (\bibinfo {year} 
			{2020})}\BibitemShut
	{NoStop}%
	\bibitem [{\citenamefont {Kannan}\ \emph {et~al.}(2023)\citenamefont 
	{Kannan},
		\citenamefont {Almanakly}, \citenamefont {Sung}, \citenamefont 
		{Di~Paolo},
		\citenamefont {Rower}, \citenamefont {Braum\"{u}ller}, \citenamefont
		{Melville}, \citenamefont {Niedzielski}, \citenamefont {Karamlou},
		\citenamefont {Serniak}, \citenamefont {Veps\"{a}l\"{a}inen}, 
		\citenamefont
		{Schwartz}, \citenamefont {Yoder}, \citenamefont {Winik}, 
		\citenamefont
		{Wang}, \citenamefont {Orlando}, \citenamefont {Gustavsson}, 
		\citenamefont
		{Grover},\ and\ \citenamefont {Oliver}}]{Kannan2023}%
	\BibitemOpen
	\bibfield  {author} {\bibinfo {author} {\bibfnamefont {B.}~\bibnamefont
			{Kannan}}, \bibinfo {author} {\bibfnamefont {A.}~\bibnamefont 
			{Almanakly}},
		\bibinfo {author} {\bibfnamefont {Y.}~\bibnamefont {Sung}}, \bibinfo 
		{author}
		{\bibfnamefont {A.}~\bibnamefont {Di~Paolo}}, \bibinfo {author}
		{\bibfnamefont {D.~A.}\ \bibnamefont {Rower}}, \bibinfo {author}
		{\bibfnamefont {J.}~\bibnamefont {Braum\"{u}ller}}, \bibinfo {author}
		{\bibfnamefont {A.}~\bibnamefont {Melville}}, \bibinfo {author}
		{\bibfnamefont {B.~M.}\ \bibnamefont {Niedzielski}}, \bibinfo {author}
		{\bibfnamefont {A.}~\bibnamefont {Karamlou}}, \bibinfo {author}
		{\bibfnamefont {K.}~\bibnamefont {Serniak}}, \bibinfo {author} 
		{\bibfnamefont
			{A.}~\bibnamefont {Veps\"{a}l\"{a}inen}}, \bibinfo {author} 
			{\bibfnamefont
			{M.~E.}\ \bibnamefont {Schwartz}}, \bibinfo {author} 
			{\bibfnamefont {J.~L.}\
			\bibnamefont {Yoder}}, \bibinfo {author} {\bibfnamefont 
			{R.}~\bibnamefont
			{Winik}}, \bibinfo {author} {\bibfnamefont {J.~I.-J.}\ 
			\bibnamefont {Wang}},
		\bibinfo {author} {\bibfnamefont {T.~P.}\ \bibnamefont {Orlando}}, 
		\bibinfo
		{author} {\bibfnamefont {S.}~\bibnamefont {Gustavsson}}, \bibinfo 
		{author}
		{\bibfnamefont {J.~A.}\ \bibnamefont {Grover}},\ and\ \bibinfo 
		{author}
		{\bibfnamefont {W.~D.}\ \bibnamefont {Oliver}},\ }\bibfield  {title}
	{\bibinfo {title} {On-demand directional microwave photon emission using
			waveguide quantum electrodynamics},\ }\href
	{https://doi.org/10.1038/s41567-022-01869-5} {\bibfield  {journal} 
	{\bibinfo
			{journal} {Nature Physics}\ }\textbf {\bibinfo {volume} {19}},\ 
			\bibinfo
		{pages} {394} (\bibinfo {year} {2023})}\BibitemShut {NoStop}%
	\bibitem [{\citenamefont {Yao}\ \emph {et~al.}(2005)\citenamefont {Yao},
		\citenamefont {Liu},\ and\ \citenamefont {Sham}}]{Yao2005}%
	\BibitemOpen
	\bibfield  {author} {\bibinfo {author} {\bibfnamefont {W.}~\bibnamefont
			{Yao}}, \bibinfo {author} {\bibfnamefont {R.-B.}\ \bibnamefont 
			{Liu}},\ and\
		\bibinfo {author} {\bibfnamefont {L.~J.}\ \bibnamefont {Sham}},\ 
		}\bibfield
	{title} {\bibinfo {title} {Theory of control of the spin-photon interface 
	for
			quantum networks},\ }\href 
			{https://doi.org/10.1103/PhysRevLett.95.030504}
	{\bibfield  {journal} {\bibinfo  {journal} {Phys. Rev. Lett.}\ }\textbf
		{\bibinfo {volume} {95}},\ \bibinfo {pages} {030504} (\bibinfo {year}
		{2005})}\BibitemShut {NoStop}%
	\bibitem [{\citenamefont {Stannigel}\ \emph {et~al.}(2011)\citenamefont
		{Stannigel}, \citenamefont {Rabl}, \citenamefont {S\o{}rensen}, 
		\citenamefont
		{Lukin},\ and\ \citenamefont {Zoller}}]{Stannigel2011}%
	\BibitemOpen
	\bibfield  {author} {\bibinfo {author} {\bibfnamefont {K.}~\bibnamefont
			{Stannigel}}, \bibinfo {author} {\bibfnamefont {P.}~\bibnamefont 
			{Rabl}},
		\bibinfo {author} {\bibfnamefont {A.~S.}\ \bibnamefont {S\o{}rensen}},
		\bibinfo {author} {\bibfnamefont {M.~D.}\ \bibnamefont {Lukin}},\ and\
		\bibinfo {author} {\bibfnamefont {P.}~\bibnamefont {Zoller}},\ 
		}\bibfield
	{title} {\bibinfo {title} {Optomechanical transducers for 
	quantum-information
			processing},\ }\href {https://doi.org/10.1103/PhysRevA.84.042341} 
			{\bibfield
		{journal} {\bibinfo  {journal} {Phys. Rev. A}\ }\textbf {\bibinfo 
		{volume}
			{84}},\ \bibinfo {pages} {042341} (\bibinfo {year} 
			{2011})}\BibitemShut
	{NoStop}%
	\bibitem [{\citenamefont {Stannigel}\ \emph {et~al.}(2012)\citenamefont
		{Stannigel}, \citenamefont {Rabl},\ and\ \citenamefont
		{Zoller}}]{Stannigel2012}%
	\BibitemOpen
	\bibfield  {author} {\bibinfo {author} {\bibfnamefont {K.}~\bibnamefont
			{Stannigel}}, \bibinfo {author} {\bibfnamefont {P.}~\bibnamefont 
			{Rabl}},\
		and\ \bibinfo {author} {\bibfnamefont {P.}~\bibnamefont {Zoller}},\
	}\bibfield  {title} {\bibinfo {title} {Driven-dissipative preparation of
			entangled states in cascaded quantum-optical networks},\ }\href
	{https://doi.org/10.1088/1367-2630/14/6/063014} {\bibfield  {journal}
		{\bibinfo  {journal} {New Journal of Physics}\ }\textbf {\bibinfo 
		{volume}
			{14}},\ \bibinfo {pages} {063014} (\bibinfo {year} 
			{2012})}\BibitemShut
	{NoStop}%
	\bibitem [{\citenamefont {Vermersch}\ \emph {et~al.}(2017)\citenamefont
		{Vermersch}, \citenamefont {Guimond}, \citenamefont {Pichler},\ and\
		\citenamefont {Zoller}}]{Vermersch2017}%
	\BibitemOpen
	\bibfield  {author} {\bibinfo {author} {\bibfnamefont {B.}~\bibnamefont
			{Vermersch}}, \bibinfo {author} {\bibfnamefont {P.-O.}\ 
			\bibnamefont
			{Guimond}}, \bibinfo {author} {\bibfnamefont {H.}~\bibnamefont 
			{Pichler}},\
		and\ \bibinfo {author} {\bibfnamefont {P.}~\bibnamefont {Zoller}},\
	}\bibfield  {title} {\bibinfo {title} {{Quantum State Transfer via Noisy
				Photonic and Phononic Waveguides}},\ }\href
	{https://doi.org/10.1103/PhysRevLett.118.133601} {\bibfield  {journal}
		{\bibinfo  {journal} {Phys. Rev. Lett.}\ }\textbf {\bibinfo {volume} 
		{118}},\
		\bibinfo {pages} {133601} (\bibinfo {year} {2017})}\BibitemShut 
		{NoStop}%
	\bibitem [{\citenamefont {Xiang}\ \emph {et~al.}(2017)\citenamefont 
	{Xiang},
		\citenamefont {Zhang}, \citenamefont {Jiang},\ and\ \citenamefont
		{Rabl}}]{Xiang2017}%
	\BibitemOpen
	\bibfield  {author} {\bibinfo {author} {\bibfnamefont {Z.-L.}\ 
	\bibnamefont
			{Xiang}}, \bibinfo {author} {\bibfnamefont {M.}~\bibnamefont 
			{Zhang}},
		\bibinfo {author} {\bibfnamefont {L.}~\bibnamefont {Jiang}},\ and\ 
		\bibinfo
		{author} {\bibfnamefont {P.}~\bibnamefont {Rabl}},\ }\bibfield  
		{title}
	{\bibinfo {title} {{Intracity Quantum Communication via Thermal Microwave
				Networks}},\ }\href 
				{https://doi.org/10.1103/PhysRevX.7.011035} {\bibfield
		{journal} {\bibinfo  {journal} {Phys. Rev. X}\ }\textbf {\bibinfo 
		{volume}
			{7}},\ \bibinfo {pages} {011035} (\bibinfo {year} 
			{2017})}\BibitemShut
	{NoStop}%
	\bibitem [{\citenamefont {Chang}\ \emph {et~al.}(2014)\citenamefont 
	{Chang},
		\citenamefont {Vuletic},\ and\ \citenamefont {Lukin}}]{Chang2014}%
	\BibitemOpen
	\bibfield  {author} {\bibinfo {author} {\bibfnamefont {D.~E.}\ 
	\bibnamefont
			{Chang}}, \bibinfo {author} {\bibfnamefont {V.}~\bibnamefont 
			{Vuletic}},\
		and\ \bibinfo {author} {\bibfnamefont {M.~D.}\ \bibnamefont {Lukin}},\
	}\bibfield  {title} {\bibinfo {title} {Quantum nonlinear optics -- photon 
	by
			photon},\ }\href {https://doi.org/10.1038/nphoton.2014.192} 
			{\bibfield
		{journal} {\bibinfo  {journal} {Nature Photonics}\ }\textbf {\bibinfo
			{volume} {8}},\ \bibinfo {pages} {685} (\bibinfo {year} 
			{2014})}\BibitemShut
	{NoStop}%
	\bibitem [{\citenamefont {Mahmoodian}\ \emph {et~al.}(2018)\citenamefont
		{Mahmoodian}, \citenamefont {\ifmmode~\check{C}\else 
		\v{C}\fi{}epulkovskis},
		\citenamefont {Das}, \citenamefont {Lodahl}, \citenamefont 
		{Hammerer},\ and\
		\citenamefont {S\o{}rensen}}]{Mahmoodian2018}%
	\BibitemOpen
	\bibfield  {author} {\bibinfo {author} {\bibfnamefont {S.}~\bibnamefont
			{Mahmoodian}}, \bibinfo {author} {\bibfnamefont {M.}~\bibnamefont
			{\ifmmode~\check{C}\else \v{C}\fi{}epulkovskis}}, \bibinfo 
			{author}
		{\bibfnamefont {S.}~\bibnamefont {Das}}, \bibinfo {author} 
		{\bibfnamefont
			{P.}~\bibnamefont {Lodahl}}, \bibinfo {author} {\bibfnamefont
			{K.}~\bibnamefont {Hammerer}},\ and\ \bibinfo {author} 
			{\bibfnamefont
			{A.~S.}\ \bibnamefont {S\o{}rensen}},\ }\bibfield  {title} 
			{\bibinfo {title}
		{{Strongly Correlated Photon Transport in Waveguide Quantum 
		Electrodynamics
				with Weakly Coupled Emitters}},\ }\href
	{https://doi.org/10.1103/PhysRevLett.121.143601} {\bibfield  {journal}
		{\bibinfo  {journal} {Phys. Rev. Lett.}\ }\textbf {\bibinfo {volume} 
		{121}},\
		\bibinfo {pages} {143601} (\bibinfo {year} {2018})}\BibitemShut 
		{NoStop}%
	\bibitem [{\citenamefont {Solano}\ \emph {et~al.}(2023)\citenamefont 
	{Solano},
		\citenamefont {Barberis-Blostein},\ and\ \citenamefont 
		{Sinha}}]{Solano2023}%
	\BibitemOpen
	\bibfield  {author} {\bibinfo {author} {\bibfnamefont {P.}~\bibnamefont
			{Solano}}, \bibinfo {author} {\bibfnamefont {P.}~\bibnamefont
			{Barberis-Blostein}},\ and\ \bibinfo {author} {\bibfnamefont
			{K.}~\bibnamefont {Sinha}},\ }\bibfield  {title} {\bibinfo {title}
		{Dissimilar collective decay and directional emission from two quantum
			emitters},\ }\href {https://doi.org/10.1103/PhysRevA.107.023723} 
			{\bibfield
		{journal} {\bibinfo  {journal} {Phys. Rev. A}\ }\textbf {\bibinfo 
		{volume}
			{107}},\ \bibinfo {pages} {023723} (\bibinfo {year} 
			{2023})}\BibitemShut
	{NoStop}%
	\bibitem [{\citenamefont {Nagata}\ \emph {et~al.}(2007)\citenamefont 
	{Nagata},
		\citenamefont {Okamoto}, \citenamefont {O'Brien}, \citenamefont 
		{Sasaki},\
		and\ \citenamefont {Takeuchi}}]{Nagata2007}%
	\BibitemOpen
	\bibfield  {author} {\bibinfo {author} {\bibfnamefont {T.}~\bibnamefont
			{Nagata}}, \bibinfo {author} {\bibfnamefont {R.}~\bibnamefont 
			{Okamoto}},
		\bibinfo {author} {\bibfnamefont {J.~L.}\ \bibnamefont {O'Brien}}, 
		\bibinfo
		{author} {\bibfnamefont {K.}~\bibnamefont {Sasaki}},\ and\ \bibinfo 
		{author}
		{\bibfnamefont {S.}~\bibnamefont {Takeuchi}},\ }\bibfield  {title} 
		{\bibinfo
		{title} {{Beating the Standard Quantum Limit with Four-Entangled 
		Photons}},\
	}\href {https://doi.org/10.1126/science.1138007} {\bibfield  {journal}
		{\bibinfo  {journal} {Science}\ }\textbf {\bibinfo {volume} {316}},\ 
		\bibinfo
		{pages} {726} (\bibinfo {year} {2007})}\BibitemShut {NoStop}%
	\bibitem [{\citenamefont {Yuan}\ \emph {et~al.}(2010)\citenamefont {Yuan},
		\citenamefont {Bao}, \citenamefont {Lu}, \citenamefont {Zhang}, 
		\citenamefont
		{Peng},\ and\ \citenamefont {Pan}}]{YUAN20101}%
	\BibitemOpen
	\bibfield  {author} {\bibinfo {author} {\bibfnamefont {Z.-S.}\ 
	\bibnamefont
			{Yuan}}, \bibinfo {author} {\bibfnamefont {X.-H.}\ \bibnamefont 
			{Bao}},
		\bibinfo {author} {\bibfnamefont {C.-Y.}\ \bibnamefont {Lu}}, \bibinfo
		{author} {\bibfnamefont {J.}~\bibnamefont {Zhang}}, \bibinfo {author}
		{\bibfnamefont {C.-Z.}\ \bibnamefont {Peng}},\ and\ \bibinfo {author}
		{\bibfnamefont {J.-W.}\ \bibnamefont {Pan}},\ }\bibfield  {title} 
		{\bibinfo
		{title} {Entangled photons and quantum communication},\ }\href
	{https://doi.org/https://doi.org/10.1016/j.physrep.2010.07.004} {\bibfield
		{journal} {\bibinfo  {journal} {Physics Reports}\ }\textbf {\bibinfo 
		{volume}
			{497}},\ \bibinfo {pages} {1} (\bibinfo {year} 
			{2010})}\BibitemShut {NoStop}%
	\bibitem [{\citenamefont {Giovannetti}\ \emph {et~al.}(2011)\citenamefont
		{Giovannetti}, \citenamefont {Lloyd},\ and\ \citenamefont
		{Maccone}}]{Giovannetti2011}%
	\BibitemOpen
	\bibfield  {author} {\bibinfo {author} {\bibfnamefont {V.}~\bibnamefont
			{Giovannetti}}, \bibinfo {author} {\bibfnamefont 
			{S.}~\bibnamefont {Lloyd}},\
		and\ \bibinfo {author} {\bibfnamefont {L.}~\bibnamefont {Maccone}},\
	}\bibfield  {title} {\bibinfo {title} {Advances in quantum metrology},\
	}\href {https://doi.org/10.1038/nphoton.2011.35} {\bibfield  {journal}
		{\bibinfo  {journal} {Nature Photonics}\ }\textbf {\bibinfo {volume} 
		{5}},\
		\bibinfo {pages} {222} (\bibinfo {year} {2011})}\BibitemShut {NoStop}%
	\bibitem [{\citenamefont {Gatto~Monticone}\ \emph 
	{et~al.}(2014)\citenamefont
		{Gatto~Monticone}, \citenamefont {Katamadze}, \citenamefont {Traina},
		\citenamefont {Moreva}, \citenamefont {Forneris}, \citenamefont
		{Ruo-Berchera}, \citenamefont {Olivero}, \citenamefont {Degiovanni},
		\citenamefont {Brida},\ and\ \citenamefont 
		{Genovese}}]{Monticone2014}%
	\BibitemOpen
	\bibfield  {author} {\bibinfo {author} {\bibfnamefont {D.}~\bibnamefont
			{Gatto~Monticone}}, \bibinfo {author} {\bibfnamefont 
			{K.}~\bibnamefont
			{Katamadze}}, \bibinfo {author} {\bibfnamefont {P.}~\bibnamefont 
			{Traina}},
		\bibinfo {author} {\bibfnamefont {E.}~\bibnamefont {Moreva}}, \bibinfo
		{author} {\bibfnamefont {J.}~\bibnamefont {Forneris}}, \bibinfo 
		{author}
		{\bibfnamefont {I.}~\bibnamefont {Ruo-Berchera}}, \bibinfo {author}
		{\bibfnamefont {P.}~\bibnamefont {Olivero}}, \bibinfo {author} 
		{\bibfnamefont
			{I.~P.}\ \bibnamefont {Degiovanni}}, \bibinfo {author} 
			{\bibfnamefont
			{G.}~\bibnamefont {Brida}},\ and\ \bibinfo {author} {\bibfnamefont
			{M.}~\bibnamefont {Genovese}},\ }\bibfield  {title} {\bibinfo 
			{title}
		{{Beating the Abbe Diffraction Limit in Confocal Microscopy via 
		Nonclassical
				Photon Statistics}},\ }\href 
				{https://doi.org/10.1103/PhysRevLett.113.143602}
	{\bibfield  {journal} {\bibinfo  {journal} {Phys. Rev. Lett.}\ }\textbf
		{\bibinfo {volume} {113}},\ \bibinfo {pages} {143602} (\bibinfo {year}
		{2014})}\BibitemShut {NoStop}%
	\bibitem [{\citenamefont {Paulisch}\ \emph {et~al.}(2019)\citenamefont
		{Paulisch}, \citenamefont {Perarnau-Llobet}, \citenamefont
		{Gonz\'alez-Tudela},\ and\ \citenamefont {Cirac}}]{Paulisch2019}%
	\BibitemOpen
	\bibfield  {author} {\bibinfo {author} {\bibfnamefont {V.}~\bibnamefont
			{Paulisch}}, \bibinfo {author} {\bibfnamefont {M.}~\bibnamefont
			{Perarnau-Llobet}}, \bibinfo {author} {\bibfnamefont 
			{A.}~\bibnamefont
			{Gonz\'alez-Tudela}},\ and\ \bibinfo {author} {\bibfnamefont 
			{J.~I.}\
			\bibnamefont {Cirac}},\ }\bibfield  {title} {\bibinfo {title} 
			{Quantum
			metrology with one-dimensional superradiant photonic states},\ 
			}\href
	{https://doi.org/10.1103/PhysRevA.99.043807} {\bibfield  {journal} 
	{\bibinfo
			{journal} {Phys. Rev. A}\ }\textbf {\bibinfo {volume} {99}},\ 
			\bibinfo
		{pages} {043807} (\bibinfo {year} {2019})}\BibitemShut {NoStop}%
	\bibitem [{\citenamefont {Sheremet}\ \emph {et~al.}(2023)\citenamefont
		{Sheremet}, \citenamefont {Petrov}, \citenamefont {Iorsh}, 
		\citenamefont
		{Poshakinskiy},\ and\ \citenamefont {Poddubny}}]{Sheremet2023}%
	\BibitemOpen
	\bibfield  {author} {\bibinfo {author} {\bibfnamefont {A.~S.}\ 
	\bibnamefont
			{Sheremet}}, \bibinfo {author} {\bibfnamefont {M.~I.}\ 
			\bibnamefont
			{Petrov}}, \bibinfo {author} {\bibfnamefont {I.~V.}\ \bibnamefont 
			{Iorsh}},
		\bibinfo {author} {\bibfnamefont {A.~V.}\ \bibnamefont 
		{Poshakinskiy}},\ and\
		\bibinfo {author} {\bibfnamefont {A.~N.}\ \bibnamefont {Poddubny}},\
	}\bibfield  {title} {\bibinfo {title} {Waveguide quantum electrodynamics:
			Collective radiance and photon-photon correlations},\ }\href
	{https://doi.org/10.1103/RevModPhys.95.015002} {\bibfield  {journal}
		{\bibinfo  {journal} {Rev. Mod. Phys.}\ }\textbf {\bibinfo {volume} 
		{95}},\
		\bibinfo {pages} {015002} (\bibinfo {year} {2023})}\BibitemShut 
		{NoStop}%
	\bibitem [{\citenamefont {Mahmoodian}\ \emph {et~al.}(2020)\citenamefont
		{Mahmoodian}, \citenamefont {Calaj\'o}, \citenamefont {Chang}, 
		\citenamefont
		{Hammerer},\ and\ \citenamefont {S\o{}rensen}}]{Mahmoodian2020}%
	\BibitemOpen
	\bibfield  {author} {\bibinfo {author} {\bibfnamefont {S.}~\bibnamefont
			{Mahmoodian}}, \bibinfo {author} {\bibfnamefont {G.}~\bibnamefont
			{Calaj\'o}}, \bibinfo {author} {\bibfnamefont {D.~E.}\ 
			\bibnamefont {Chang}},
		\bibinfo {author} {\bibfnamefont {K.}~\bibnamefont {Hammerer}},\ and\
		\bibinfo {author} {\bibfnamefont {A.~S.}\ \bibnamefont 
		{S\o{}rensen}},\
	}\bibfield  {title} {\bibinfo {title} {{Dynamics of Many-Body Photon Bound
				States in Chiral Waveguide QED}},\ }\href
	{https://doi.org/10.1103/PhysRevX.10.031011} {\bibfield  {journal} 
	{\bibinfo
			{journal} {Phys. Rev. X}\ }\textbf {\bibinfo {volume} {10}},\ 
			\bibinfo
		{pages} {031011} (\bibinfo {year} {2020})}\BibitemShut {NoStop}%
	\bibitem [{\citenamefont {Peyronel}\ \emph {et~al.}(2012)\citenamefont
		{Peyronel}, \citenamefont {Firstenberg}, \citenamefont {Liang}, 
		\citenamefont
		{Hofferberth}, \citenamefont {Gorshkov}, \citenamefont {Pohl}, 
		\citenamefont
		{Lukin},\ and\ \citenamefont {Vuletic}}]{Peyronel2012}%
	\BibitemOpen
	\bibfield  {author} {\bibinfo {author} {\bibfnamefont {T.}~\bibnamefont
			{Peyronel}}, \bibinfo {author} {\bibfnamefont {O.}~\bibnamefont
			{Firstenberg}}, \bibinfo {author} {\bibfnamefont {Q.-Y.}\ 
			\bibnamefont
			{Liang}}, \bibinfo {author} {\bibfnamefont {S.}~\bibnamefont 
			{Hofferberth}},
		\bibinfo {author} {\bibfnamefont {A.~V.}\ \bibnamefont {Gorshkov}}, 
		\bibinfo
		{author} {\bibfnamefont {T.}~\bibnamefont {Pohl}}, \bibinfo {author}
		{\bibfnamefont {M.~D.}\ \bibnamefont {Lukin}},\ and\ \bibinfo {author}
		{\bibfnamefont {V.}~\bibnamefont {Vuletic}},\ }\bibfield  {title} 
		{\bibinfo
		{title} {Quantum nonlinear optics with single photons enabled by 
		strongly
			interacting atoms},\ }\href {https://doi.org/10.1038/nature11361} 
			{\bibfield
		{journal} {\bibinfo  {journal} {Nature}\ }\textbf {\bibinfo {volume} 
		{488}},\
		\bibinfo {pages} {57} (\bibinfo {year} {2012})}\BibitemShut {NoStop}%
	\bibitem [{\citenamefont {Roy}\ \emph {et~al.}(2017)\citenamefont {Roy},
		\citenamefont {Wilson},\ and\ \citenamefont 
		{Firstenberg}}]{Dibyendu2017}%
	\BibitemOpen
	\bibfield  {author} {\bibinfo {author} {\bibfnamefont {D.}~\bibnamefont
			{Roy}}, \bibinfo {author} {\bibfnamefont {C.~M.}\ \bibnamefont 
			{Wilson}},\
		and\ \bibinfo {author} {\bibfnamefont {O.}~\bibnamefont 
		{Firstenberg}},\
	}\bibfield  {title} {\bibinfo {title} {{Colloquium: Strongly interacting
				photons in one-dimensional continuum}},\ }\href
	{https://doi.org/10.1103/RevModPhys.89.021001} {\bibfield  {journal}
		{\bibinfo  {journal} {Rev. Mod. Phys.}\ }\textbf {\bibinfo {volume} 
		{89}},\
		\bibinfo {pages} {021001} (\bibinfo {year} {2017})}\BibitemShut 
		{NoStop}%
	\bibitem [{\citenamefont {Koch}\ \emph {et~al.}(2007)\citenamefont {Koch},
		\citenamefont {Yu}, \citenamefont {Gambetta}, \citenamefont {Houck},
		\citenamefont {Schuster}, \citenamefont {Majer}, \citenamefont 
		{Blais},
		\citenamefont {Devoret}, \citenamefont {Girvin},\ and\ \citenamefont
		{Schoelkopf}}]{Koch2007}%
	\BibitemOpen
	\bibfield  {author} {\bibinfo {author} {\bibfnamefont {J.}~\bibnamefont
			{Koch}}, \bibinfo {author} {\bibfnamefont {T.~M.}\ \bibnamefont 
			{Yu}},
		\bibinfo {author} {\bibfnamefont {J.}~\bibnamefont {Gambetta}}, 
		\bibinfo
		{author} {\bibfnamefont {A.~A.}\ \bibnamefont {Houck}}, \bibinfo 
		{author}
		{\bibfnamefont {D.~I.}\ \bibnamefont {Schuster}}, \bibinfo {author}
		{\bibfnamefont {J.}~\bibnamefont {Majer}}, \bibinfo {author} 
		{\bibfnamefont
			{A.}~\bibnamefont {Blais}}, \bibinfo {author} {\bibfnamefont 
			{M.~H.}\
			\bibnamefont {Devoret}}, \bibinfo {author} {\bibfnamefont {S.~M.}\
			\bibnamefont {Girvin}},\ and\ \bibinfo {author} {\bibfnamefont 
			{R.~J.}\
			\bibnamefont {Schoelkopf}},\ }\bibfield  {title} {\bibinfo {title}
		{{Charge-insensitive qubit design derived from the Cooper pair 
		box}},\ }\href
	{https://doi.org/10.1103/PhysRevA.76.042319} {\bibfield  {journal} 
	{\bibinfo
			{journal} {Phys. Rev. A}\ }\textbf {\bibinfo {volume} {76}},\ 
			\bibinfo
		{pages} {042319} (\bibinfo {year} {2007})}\BibitemShut {NoStop}%
	\bibitem [{\citenamefont {Orell}\ \emph {et~al.}(2019)\citenamefont 
	{Orell},
		\citenamefont {Michailidis}, \citenamefont {Serbyn},\ and\ 
		\citenamefont
		{Silveri}}]{Orell2019}%
	\BibitemOpen
	\bibfield  {author} {\bibinfo {author} {\bibfnamefont {T.}~\bibnamefont
			{Orell}}, \bibinfo {author} {\bibfnamefont {A.~A.}\ \bibnamefont
			{Michailidis}}, \bibinfo {author} {\bibfnamefont {M.}~\bibnamefont
			{Serbyn}},\ and\ \bibinfo {author} {\bibfnamefont 
			{M.}~\bibnamefont
			{Silveri}},\ }\bibfield  {title} {\bibinfo {title} {Probing the 
			many-body
			localization phase transition with superconducting circuits},\ 
			}\href
	{https://doi.org/10.1103/PhysRevB.100.134504} {\bibfield  {journal} 
	{\bibinfo
			{journal} {Phys. Rev. B}\ }\textbf {\bibinfo {volume} {100}},\ 
			\bibinfo
		{pages} {134504} (\bibinfo {year} {2019})}\BibitemShut {NoStop}%
	\bibitem [{\citenamefont {Blais}\ \emph {et~al.}(2021)\citenamefont 
	{Blais},
		\citenamefont {Grimsmo}, \citenamefont {Girvin},\ and\ \citenamefont
		{Wallraff}}]{Blais2021}%
	\BibitemOpen
	\bibfield  {author} {\bibinfo {author} {\bibfnamefont {A.}~\bibnamefont
			{Blais}}, \bibinfo {author} {\bibfnamefont {A.~L.}\ \bibnamefont 
			{Grimsmo}},
		\bibinfo {author} {\bibfnamefont {S.~M.}\ \bibnamefont {Girvin}},\ 
		and\
		\bibinfo {author} {\bibfnamefont {A.}~\bibnamefont {Wallraff}},\ 
		}\bibfield
	{title} {\bibinfo {title} {{Circuit quantum electrodynamics}},\ }\href
	{https://doi.org/10.1103/RevModPhys.93.025005} {\bibfield  {journal}
		{\bibinfo  {journal} {Rev. Mod. Phys.}\ }\textbf {\bibinfo {volume} 
		{93}},\
		\bibinfo {pages} {025005} (\bibinfo {year} {2021})}\BibitemShut 
		{NoStop}%
	\bibitem [{\citenamefont {Lukin}\ \emph {et~al.}(2001)\citenamefont 
	{Lukin},
		\citenamefont {Fleischhauer}, \citenamefont {Cote}, \citenamefont 
		{Duan},
		\citenamefont {Jaksch}, \citenamefont {Cirac},\ and\ \citenamefont
		{Zoller}}]{Lukin2001}%
	\BibitemOpen
	\bibfield  {author} {\bibinfo {author} {\bibfnamefont {M.~D.}\ 
	\bibnamefont
			{Lukin}}, \bibinfo {author} {\bibfnamefont {M.}~\bibnamefont 
			{Fleischhauer}},
		\bibinfo {author} {\bibfnamefont {R.}~\bibnamefont {Cote}}, \bibinfo 
		{author}
		{\bibfnamefont {L.~M.}\ \bibnamefont {Duan}}, \bibinfo {author}
		{\bibfnamefont {D.}~\bibnamefont {Jaksch}}, \bibinfo {author} 
		{\bibfnamefont
			{J.~I.}\ \bibnamefont {Cirac}},\ and\ \bibinfo {author} 
			{\bibfnamefont
			{P.}~\bibnamefont {Zoller}},\ }\bibfield  {title} {\bibinfo 
			{title} {{Dipole
				Blockade and Quantum Information Processing in Mesoscopic 
				Atomic
				Ensembles}},\ }\href 
				{https://doi.org/10.1103/PhysRevLett.87.037901}
	{\bibfield  {journal} {\bibinfo  {journal} {Phys. Rev. Lett.}\ }\textbf
		{\bibinfo {volume} {87}},\ \bibinfo {pages} {037901} (\bibinfo {year}
		{2001})}\BibitemShut {NoStop}%
	\bibitem [{\citenamefont {Gorshkov}\ \emph {et~al.}(2011)\citenamefont
		{Gorshkov}, \citenamefont {Otterbach}, \citenamefont {Fleischhauer},
		\citenamefont {Pohl},\ and\ \citenamefont {Lukin}}]{Gorshkov2011}%
	\BibitemOpen
	\bibfield  {author} {\bibinfo {author} {\bibfnamefont {A.~V.}\ 
	\bibnamefont
			{Gorshkov}}, \bibinfo {author} {\bibfnamefont {J.}~\bibnamefont 
			{Otterbach}},
		\bibinfo {author} {\bibfnamefont {M.}~\bibnamefont {Fleischhauer}}, 
		\bibinfo
		{author} {\bibfnamefont {T.}~\bibnamefont {Pohl}},\ and\ \bibinfo 
		{author}
		{\bibfnamefont {M.~D.}\ \bibnamefont {Lukin}},\ }\bibfield  {title} 
		{\bibinfo
		{title} {{Photon-Photon Interactions via Rydberg Blockade}},\ }\href
	{https://doi.org/10.1103/PhysRevLett.107.133602} {\bibfield  {journal}
		{\bibinfo  {journal} {Phys. Rev. Lett.}\ }\textbf {\bibinfo {volume} 
		{107}},\
		\bibinfo {pages} {133602} (\bibinfo {year} {2011})}\BibitemShut 
		{NoStop}%
	\bibitem [{\citenamefont {Fedorov}\ \emph {et~al.}(2021)\citenamefont
		{Fedorov}, \citenamefont {Remizov}, \citenamefont {Shapiro}, 
		\citenamefont
		{Pogosov}, \citenamefont {Egorova}, \citenamefont {Tsitsilin}, 
		\citenamefont
		{Andronik}, \citenamefont {Dobronosova}, \citenamefont {Rodionov},
		\citenamefont {Astafiev},\ and\ \citenamefont 
		{Ustinov}}]{Fedorov2021}%
	\BibitemOpen
	\bibfield  {author} {\bibinfo {author} {\bibfnamefont {G.~P.}\ 
	\bibnamefont
			{Fedorov}}, \bibinfo {author} {\bibfnamefont {S.~V.}\ \bibnamefont
			{Remizov}}, \bibinfo {author} {\bibfnamefont {D.~S.}\ \bibnamefont
			{Shapiro}}, \bibinfo {author} {\bibfnamefont {W.~V.}\ \bibnamefont
			{Pogosov}}, \bibinfo {author} {\bibfnamefont {E.}~\bibnamefont 
			{Egorova}},
		\bibinfo {author} {\bibfnamefont {I.}~\bibnamefont {Tsitsilin}}, 
		\bibinfo
		{author} {\bibfnamefont {M.}~\bibnamefont {Andronik}}, \bibinfo 
		{author}
		{\bibfnamefont {A.~A.}\ \bibnamefont {Dobronosova}}, \bibinfo {author}
		{\bibfnamefont {I.~A.}\ \bibnamefont {Rodionov}}, \bibinfo {author}
		{\bibfnamefont {O.~V.}\ \bibnamefont {Astafiev}},\ and\ \bibinfo 
		{author}
		{\bibfnamefont {A.~V.}\ \bibnamefont {Ustinov}},\ }\bibfield  {title}
	{\bibinfo {title} {Photon transport in a bose-hubbard chain of
			superconducting artificial atoms},\ }\href
	{https://doi.org/10.1103/PhysRevLett.126.180503} {\bibfield  {journal}
		{\bibinfo  {journal} {Phys. Rev. Lett.}\ }\textbf {\bibinfo {volume} 
		{126}},\
		\bibinfo {pages} {180503} (\bibinfo {year} {2021})}\BibitemShut 
		{NoStop}%
	\bibitem [{\citenamefont {Karamlou}\ \emph {et~al.}(2024)\citenamefont
		{Karamlou}, \citenamefont {Rosen}, \citenamefont {Muschinske}, 
		\citenamefont
		{Barrett}, \citenamefont {Di~Paolo}, \citenamefont {Ding}, 
		\citenamefont
		{Harrington}, \citenamefont {Hays}, \citenamefont {Das}, 
		\citenamefont {Kim},
		\citenamefont {Niedzielski}, \citenamefont {Schuldt}, \citenamefont
		{Serniak}, \citenamefont {Schwartz}, \citenamefont {Yoder}, 
		\citenamefont
		{Gustavsson}, \citenamefont {Yanay}, \citenamefont {Grover},\ and\
		\citenamefont {Oliver}}]{Karamlou2024}%
	\BibitemOpen
	\bibfield  {author} {\bibinfo {author} {\bibfnamefont {A.~H.}\ 
	\bibnamefont
			{Karamlou}}, \bibinfo {author} {\bibfnamefont {I.~T.}\ 
			\bibnamefont {Rosen}},
		\bibinfo {author} {\bibfnamefont {S.~E.}\ \bibnamefont {Muschinske}},
		\bibinfo {author} {\bibfnamefont {C.~N.}\ \bibnamefont {Barrett}}, 
		\bibinfo
		{author} {\bibfnamefont {A.}~\bibnamefont {Di~Paolo}}, \bibinfo 
		{author}
		{\bibfnamefont {L.}~\bibnamefont {Ding}}, \bibinfo {author} 
		{\bibfnamefont
			{P.~M.}\ \bibnamefont {Harrington}}, \bibinfo {author} 
			{\bibfnamefont
			{M.}~\bibnamefont {Hays}}, \bibinfo {author} {\bibfnamefont 
			{R.}~\bibnamefont
			{Das}}, \bibinfo {author} {\bibfnamefont {D.~K.}\ \bibnamefont 
			{Kim}},
		\bibinfo {author} {\bibfnamefont {B.~M.}\ \bibnamefont {Niedzielski}},
		\bibinfo {author} {\bibfnamefont {M.}~\bibnamefont {Schuldt}}, 
		\bibinfo
		{author} {\bibfnamefont {K.}~\bibnamefont {Serniak}}, \bibinfo 
		{author}
		{\bibfnamefont {M.~E.}\ \bibnamefont {Schwartz}}, \bibinfo {author}
		{\bibfnamefont {J.~L.}\ \bibnamefont {Yoder}}, \bibinfo {author}
		{\bibfnamefont {S.}~\bibnamefont {Gustavsson}}, \bibinfo {author}
		{\bibfnamefont {Y.}~\bibnamefont {Yanay}}, \bibinfo {author} 
		{\bibfnamefont
			{J.~A.}\ \bibnamefont {Grover}},\ and\ \bibinfo {author} 
			{\bibfnamefont
			{W.~D.}\ \bibnamefont {Oliver}},\ }\bibfield  {title} {\bibinfo 
			{title}
		{{Probing entanglement in a 2D hard-core Bose--Hubbard lattice}},\ 
		}\href
	{https://doi.org/10.1038/s41586-024-07325-z} {\bibfield  {journal} 
	{\bibinfo
			{journal} {Nature}\ }\textbf {\bibinfo {volume} {629}},\ \bibinfo 
			{pages}
		{561} (\bibinfo {year} {2024})}\BibitemShut {NoStop}%
	\bibitem [{\citenamefont {Castillo-Moreno}\ \emph 
	{et~al.}(2025)\citenamefont
		{Castillo-Moreno}, \citenamefont {S'epulcre}, \citenamefont 
		{Hillmann},
		\citenamefont {Amin}, \citenamefont {Kervinen},\ and\ \citenamefont
		{Gasparinetti}}]{Claudia2025}%
	\BibitemOpen
	\bibfield  {author} {\bibinfo {author} {\bibfnamefont {C.}~\bibnamefont
			{Castillo-Moreno}}, \bibinfo {author} {\bibfnamefont 
			{T.}~\bibnamefont
			{S'epulcre}}, \bibinfo {author} {\bibfnamefont {T.}~\bibnamefont 
			{Hillmann}},
		\bibinfo {author} {\bibfnamefont {K.~R.}\ \bibnamefont {Amin}}, 
		\bibinfo
		{author} {\bibfnamefont {M.}~\bibnamefont {Kervinen}},\ and\ \bibinfo
		{author} {\bibfnamefont {S.}~\bibnamefont {Gasparinetti}},\ 
		}\href@noop {}
	{\bibinfo {title} {Experimental observation of multimode quantum phase
			transitions in a superconducting bose-hubbard simulator}} 
			(\bibinfo {year}
	{2025}),\ \Eprint {https://arxiv.org/abs/2508.20116} {arXiv:2508.20116}
	\BibitemShut {NoStop}%
	\bibitem [{\citenamefont {Weckesser}\ \emph {et~al.}(2025)\citenamefont
		{Weckesser}, \citenamefont {Srakaew}, \citenamefont {Blatz}, 
		\citenamefont
		{Wei}, \citenamefont {Adler}, \citenamefont {Agrawal}, \citenamefont
		{Bohrdt}, \citenamefont {Bloch},\ and\ \citenamefont
		{Zeiher}}]{Weckesser2025}%
	\BibitemOpen
	\bibfield  {author} {\bibinfo {author} {\bibfnamefont {P.}~\bibnamefont
			{Weckesser}}, \bibinfo {author} {\bibfnamefont {K.}~\bibnamefont 
			{Srakaew}},
		\bibinfo {author} {\bibfnamefont {T.}~\bibnamefont {Blatz}}, \bibinfo
		{author} {\bibfnamefont {D.}~\bibnamefont {Wei}}, \bibinfo {author}
		{\bibfnamefont {D.}~\bibnamefont {Adler}}, \bibinfo {author} 
		{\bibfnamefont
			{S.}~\bibnamefont {Agrawal}}, \bibinfo {author} {\bibfnamefont
			{A.}~\bibnamefont {Bohrdt}}, \bibinfo {author} {\bibfnamefont
			{I.}~\bibnamefont {Bloch}},\ and\ \bibinfo {author} {\bibfnamefont
			{J.}~\bibnamefont {Zeiher}},\ }\bibfield  {title} {\bibinfo 
			{title}
		{Realization of a rydberg-dressed extended bose-hubbard model},\ 
		}\href
	{https://doi.org/doi:10.1126/science.adq7082} {\bibfield  {journal} 
	{\bibinfo
			{journal} {Science}\ }\textbf {\bibinfo {volume} {390}},\ 
			\bibinfo {pages}
		{849} (\bibinfo {year} {2025})}\BibitemShut {NoStop}%
	\bibitem [{\citenamefont {Winkler}\ \emph {et~al.}(2006)\citenamefont
		{Winkler}, \citenamefont {Thalhammer}, \citenamefont {Lang}, 
		\citenamefont
		{Grimm}, \citenamefont {Hecker~Denschlag}, \citenamefont {Daley},
		\citenamefont {Kantian}, \citenamefont {B\"{u}chler},\ and\ 
		\citenamefont
		{Zoller}}]{Winkler2006}%
	\BibitemOpen
	\bibfield  {author} {\bibinfo {author} {\bibfnamefont {K.}~\bibnamefont
			{Winkler}}, \bibinfo {author} {\bibfnamefont {G.}~\bibnamefont 
			{Thalhammer}},
		\bibinfo {author} {\bibfnamefont {F.}~\bibnamefont {Lang}}, \bibinfo 
		{author}
		{\bibfnamefont {R.}~\bibnamefont {Grimm}}, \bibinfo {author} 
		{\bibfnamefont
			{J.}~\bibnamefont {Hecker~Denschlag}}, \bibinfo {author} 
			{\bibfnamefont
			{A.~J.}\ \bibnamefont {Daley}}, \bibinfo {author} {\bibfnamefont
			{A.}~\bibnamefont {Kantian}}, \bibinfo {author} {\bibfnamefont 
			{H.~P.}\
			\bibnamefont {B\"{u}chler}},\ and\ \bibinfo {author} 
			{\bibfnamefont
			{P.}~\bibnamefont {Zoller}},\ }\bibfield  {title} {\bibinfo 
			{title}
		{Repulsively bound atom pairs in an optical lattice},\ }\href
	{https://doi.org/10.1038/nature04918} {\bibfield  {journal} {\bibinfo
			{journal} {Nature}\ }\textbf {\bibinfo {volume} {441}},\ \bibinfo 
			{pages}
		{853} (\bibinfo {year} {2006})}\BibitemShut {NoStop}%
	\bibitem [{\citenamefont {Piil}\ and\ \citenamefont
		{M\o{}lmer}(2007)}]{Piil2007}%
	\BibitemOpen
	\bibfield  {author} {\bibinfo {author} {\bibfnamefont {R.}~\bibnamefont
			{Piil}}\ and\ \bibinfo {author} {\bibfnamefont {K.}~\bibnamefont
			{M\o{}lmer}},\ }\bibfield  {title} {\bibinfo {title} {Tunneling 
			couplings in
			discrete lattices, single-particle band structure, and 
			eigenstates of
			interacting atom pairs},\ }\href 
			{https://doi.org/10.1103/PhysRevA.76.023607}
	{\bibfield  {journal} {\bibinfo  {journal} {Phys. Rev. A}\ }\textbf 
	{\bibinfo
			{volume} {76}},\ \bibinfo {pages} {023607} (\bibinfo {year}
		{2007})}\BibitemShut {NoStop}%
	\bibitem [{\citenamefont {Valiente}\ and\ \citenamefont
		{Petrosyan}(2008)}]{Valiente2008}%
	\BibitemOpen
	\bibfield  {author} {\bibinfo {author} {\bibfnamefont {M.}~\bibnamefont
			{Valiente}}\ and\ \bibinfo {author} {\bibfnamefont 
			{D.}~\bibnamefont
			{Petrosyan}},\ }\bibfield  {title} {\bibinfo {title} 
			{Two-particle states in
			the hubbard model},\ }\href 
			{https://doi.org/10.1088/0953-4075/41/16/161002}
	{\bibfield  {journal} {\bibinfo  {journal} {Journal of Physics B: Atomic,
				Molecular and Optical Physics}\ }\textbf {\bibinfo {volume} 
				{41}},\ \bibinfo
		{pages} {161002} (\bibinfo {year} {2008})}\BibitemShut {NoStop}%
	\bibitem [{\citenamefont {Mansikkam\"aki}\ \emph 
	{et~al.}(2022)\citenamefont
		{Mansikkam\"aki}, \citenamefont {Laine}, \citenamefont {Piltonen},\ 
		and\
		\citenamefont {Silveri}}]{Mansikkamaki2022}%
	\BibitemOpen
	\bibfield  {author} {\bibinfo {author} {\bibfnamefont {O.}~\bibnamefont
			{Mansikkam\"aki}}, \bibinfo {author} {\bibfnamefont 
			{S.}~\bibnamefont
			{Laine}}, \bibinfo {author} {\bibfnamefont {A.}~\bibnamefont 
			{Piltonen}},\
		and\ \bibinfo {author} {\bibfnamefont {M.}~\bibnamefont {Silveri}},\
	}\bibfield  {title} {\bibinfo {title} {Beyond hard-core bosons in transmon
			arrays},\ }\href {https://doi.org/10.1103/PRXQuantum.3.040314} 
			{\bibfield
		{journal} {\bibinfo  {journal} {PRX Quantum}\ }\textbf {\bibinfo 
		{volume}
			{3}},\ \bibinfo {pages} {040314} (\bibinfo {year} 
			{2022})}\BibitemShut
	{NoStop}%
	\bibitem [{\citenamefont {Wang}\ \emph {et~al.}(2020)\citenamefont {Wang},
		\citenamefont {Jaako}, \citenamefont {Kirton},\ and\ \citenamefont
		{Rabl}}]{WangZhihai2020}%
	\BibitemOpen
	\bibfield  {author} {\bibinfo {author} {\bibfnamefont {Z.}~\bibnamefont
			{Wang}}, \bibinfo {author} {\bibfnamefont {T.}~\bibnamefont 
			{Jaako}},
		\bibinfo {author} {\bibfnamefont {P.}~\bibnamefont {Kirton}},\ and\ 
		\bibinfo
		{author} {\bibfnamefont {P.}~\bibnamefont {Rabl}},\ }\bibfield  
		{title}
	{\bibinfo {title} {{Supercorrelated Radiance in Nonlinear Photonic
				Waveguides}},\ }\href 
				{https://doi.org/10.1103/PhysRevLett.124.213601}
	{\bibfield  {journal} {\bibinfo  {journal} {Phys. Rev. Lett.}\ }\textbf
		{\bibinfo {volume} {124}},\ \bibinfo {pages} {213601} (\bibinfo {year}
		{2020})}\BibitemShut {NoStop}%
	\bibitem [{\citenamefont {Talukdar}\ and\ \citenamefont
		{Blume}(2022)}]{Talukdar2022}%
	\BibitemOpen
	\bibfield  {author} {\bibinfo {author} {\bibfnamefont {J.}~\bibnamefont
			{Talukdar}}\ and\ \bibinfo {author} {\bibfnamefont 
			{D.}~\bibnamefont
			{Blume}},\ }\bibfield  {title} {\bibinfo {title} {{Two emitters 
			coupled to a
				bath with Kerr-like nonlinearity: Exponential decay, 
				fractional populations,
				and Rabi oscillations}},\ }\href
	{https://doi.org/10.1103/PhysRevA.105.063501} {\bibfield  {journal} 
	{\bibinfo
			{journal} {Phys. Rev. A}\ }\textbf {\bibinfo {volume} {105}},\ 
			\bibinfo
		{pages} {063501} (\bibinfo {year} {2022})}\BibitemShut {NoStop}%
	\bibitem [{\citenamefont {Li}\ and\ \citenamefont {Wang}(2025)}]{Li2025}%
	\BibitemOpen
	\bibfield  {author} {\bibinfo {author} {\bibfnamefont {J.-Q.}\ 
	\bibnamefont
			{Li}}\ and\ \bibinfo {author} {\bibfnamefont {X.}~\bibnamefont 
			{Wang}},\
	}\bibfield  {title} {\bibinfo {title} {Environmental quantum states 
	trigger
			emission in nonlinear photonics},\ }\href
	{https://doi.org/10.1038/s42005-025-02424-3} {\bibfield  {journal} 
	{\bibinfo
			{journal} {Communications Physics}\ }\textbf {\bibinfo {volume} 
			{8}},\
		\bibinfo {pages} {511} (\bibinfo {year} {2025})}\BibitemShut {NoStop}%
	\bibitem [{\citenamefont {Zhang}\ \emph {et~al.}(2025)\citenamefont 
	{Zhang},
		\citenamefont {Guo}, \citenamefont {Zhang}, \citenamefont {Wang},\ 
		and\
		\citenamefont {Wang}}]{Zhang2025}%
	\BibitemOpen
	\bibfield  {author} {\bibinfo {author} {\bibfnamefont {X.}~\bibnamefont
			{Zhang}}, \bibinfo {author} {\bibfnamefont {X.}~\bibnamefont 
			{Guo}}, \bibinfo
		{author} {\bibfnamefont {Y.}~\bibnamefont {Zhang}}, \bibinfo {author}
		{\bibfnamefont {X.}~\bibnamefont {Wang}},\ and\ \bibinfo {author}
		{\bibfnamefont {Z.}~\bibnamefont {Wang}},\ }\href@noop {} {\bibinfo 
		{title}
		{Quantum state preparation and transfer based on the bound state in 
		the
			doublon continuum}} (\bibinfo {year} {2025}),\ \Eprint
	{https://arxiv.org/abs/2512.01339} {arXiv:2512.01339} \BibitemShut 
	{NoStop}%
	\bibitem [{\citenamefont {Rieck}\ \emph {et~al.}(2025)\citenamefont 
	{Rieck},
		\citenamefont {Kockum},\ and\ \citenamefont 
		{Chen}}]{Rieck2025DoublonBS}%
	\BibitemOpen
	\bibfield  {author} {\bibinfo {author} {\bibfnamefont {W.}~\bibnamefont
			{Rieck}}, \bibinfo {author} {\bibfnamefont {A.~F.}\ \bibnamefont 
			{Kockum}},\
		and\ \bibinfo {author} {\bibfnamefont {G.}~\bibnamefont {Chen}},\ 
		}\href@noop
	{} {\bibinfo {title} {Doublon bound states in the continuum through giant
			atoms}} (\bibinfo {year} {2025}),\ \Eprint 
			{https://arxiv.org/abs/2511.18212}
	{arXiv:2511.18212} \BibitemShut {NoStop}%
	\bibitem [{\citenamefont {Frisk~Kockum}\ \emph {et~al.}(2014)\citenamefont
		{Frisk~Kockum}, \citenamefont {Delsing},\ and\ \citenamefont
		{Johansson}}]{Anton2014}%
	\BibitemOpen
	\bibfield  {author} {\bibinfo {author} {\bibfnamefont {A.}~\bibnamefont
			{Frisk~Kockum}}, \bibinfo {author} {\bibfnamefont 
			{P.}~\bibnamefont
			{Delsing}},\ and\ \bibinfo {author} {\bibfnamefont 
			{G.}~\bibnamefont
			{Johansson}},\ }\bibfield  {title} {\bibinfo {title} {{Designing
				frequency-dependent relaxation rates and Lamb shifts for a 
				giant artificial
				atom}},\ }\href {https://doi.org/10.1103/PhysRevA.90.013837} 
				{\bibfield
		{journal} {\bibinfo  {journal} {Phys. Rev. A}\ }\textbf {\bibinfo 
		{volume}
			{90}},\ \bibinfo {pages} {013837} (\bibinfo {year} 
			{2014})}\BibitemShut
	{NoStop}%
	\bibitem [{\citenamefont {Frisk~Kockum}(2020)}]{FriskKockum2020}%
	\BibitemOpen
	\bibfield  {author} {\bibinfo {author} {\bibfnamefont {A.}~\bibnamefont
			{Frisk~Kockum}},\ }\bibinfo {title} {{Quantum Optics with Giant 
			Atoms -- the
			First Five Years}},\ in\ \href 
			{https://doi.org/10.1007/978-981-15-5191-8_12}
	{\emph {\bibinfo {booktitle} {Mathematics for Industry}}}\ (\bibinfo
	{publisher} {Springer Singapore},\ \bibinfo {year} {2020})\ pp.\ \bibinfo
	{pages} {125--146}\BibitemShut {NoStop}%
	\bibitem [{\citenamefont {Du}\ \emph {et~al.}(2022)\citenamefont {Du},
		\citenamefont {Zhang}, \citenamefont {Wu}, \citenamefont {Kockum},\ 
		and\
		\citenamefont {Li}}]{DuLei2022}%
	\BibitemOpen
	\bibfield  {author} {\bibinfo {author} {\bibfnamefont {L.}~\bibnamefont
			{Du}}, \bibinfo {author} {\bibfnamefont {Y.}~\bibnamefont 
			{Zhang}}, \bibinfo
		{author} {\bibfnamefont {J.-H.}\ \bibnamefont {Wu}}, \bibinfo {author}
		{\bibfnamefont {A.~F.}\ \bibnamefont {Kockum}},\ and\ \bibinfo 
		{author}
		{\bibfnamefont {Y.}~\bibnamefont {Li}},\ }\bibfield  {title} {\bibinfo
		{title} {{Giant Atoms in a Synthetic Frequency Dimension}},\ }\href
	{https://doi.org/10.1103/PhysRevLett.128.223602} {\bibfield  {journal}
		{\bibinfo  {journal} {Phys. Rev. Lett.}\ }\textbf {\bibinfo {volume} 
		{128}},\
		\bibinfo {pages} {223602} (\bibinfo {year} {2022})}\BibitemShut 
		{NoStop}%
	\bibitem [{\citenamefont {Terradas-Brians\'o}\ \emph
		{et~al.}(2022)\citenamefont {Terradas-Brians\'o}, \citenamefont
		{Gonz\'alez-Guti\'errez}, \citenamefont {Nori}, \citenamefont
		{Mart\'{\i}n-Moreno},\ and\ \citenamefont {Zueco}}]{Terradas2022}%
	\BibitemOpen
	\bibfield  {author} {\bibinfo {author} {\bibfnamefont {S.}~\bibnamefont
			{Terradas-Brians\'o}}, \bibinfo {author} {\bibfnamefont {C.~A.}\ 
			\bibnamefont
			{Gonz\'alez-Guti\'errez}}, \bibinfo {author} {\bibfnamefont 
			{F.}~\bibnamefont
			{Nori}}, \bibinfo {author} {\bibfnamefont {L.}~\bibnamefont
			{Mart\'{\i}n-Moreno}},\ and\ \bibinfo {author} {\bibfnamefont
			{D.}~\bibnamefont {Zueco}},\ }\bibfield  {title} {\bibinfo {title}
		{{Ultrastrong waveguide QED with giant atoms}},\ }\href
	{https://doi.org/10.1103/PhysRevA.106.063717} {\bibfield  {journal} 
	{\bibinfo
			{journal} {Phys. Rev. A}\ }\textbf {\bibinfo {volume} {106}},\ 
			\bibinfo
		{pages} {063717} (\bibinfo {year} {2022})}\BibitemShut {NoStop}%
	\bibitem [{\citenamefont {Qiu}\ \emph {et~al.}(2023)\citenamefont {Qiu},
		\citenamefont {Wu},\ and\ \citenamefont {L\"{u}}}]{Qiu2023}%
	\BibitemOpen
	\bibfield  {author} {\bibinfo {author} {\bibfnamefont {Q.-Y.}\ 
	\bibnamefont
			{Qiu}}, \bibinfo {author} {\bibfnamefont {Y.}~\bibnamefont 
			{Wu}},\ and\
		\bibinfo {author} {\bibfnamefont {X.-Y.}\ \bibnamefont {L\"{u}}},\ 
		}\bibfield
	{title} {\bibinfo {title} {{Collective radiance of giant atoms in
				non-Markovian regime}},\ }\href 
				{https://doi.org/10.1007/s11433-022-1990-x}
	{\bibfield  {journal} {\bibinfo  {journal} {Sci. China Phys. Mech.}\ 
	}\textbf
		{\bibinfo {volume} {66}} (\bibinfo {year} {2023})}\BibitemShut 
		{NoStop}%
	\bibitem [{\citenamefont {Gustafsson}\ \emph {et~al.}(2014)\citenamefont
		{Gustafsson}, \citenamefont {Aref}, \citenamefont {Kockum}, 
		\citenamefont
		{Ekstr\"{o}m}, \citenamefont {Johansson},\ and\ \citenamefont
		{Delsing}}]{Gustafsson2014}%
	\BibitemOpen
	\bibfield  {author} {\bibinfo {author} {\bibfnamefont {M.~V.}\ 
	\bibnamefont
			{Gustafsson}}, \bibinfo {author} {\bibfnamefont {T.}~\bibnamefont 
			{Aref}},
		\bibinfo {author} {\bibfnamefont {A.~F.}\ \bibnamefont {Kockum}}, 
		\bibinfo
		{author} {\bibfnamefont {M.~K.}\ \bibnamefont {Ekstr\"{o}m}}, \bibinfo
		{author} {\bibfnamefont {G.}~\bibnamefont {Johansson}},\ and\ \bibinfo
		{author} {\bibfnamefont {P.}~\bibnamefont {Delsing}},\ }\bibfield  
		{title}
	{\bibinfo {title} {Propagating phonons coupled to an artificial atom},\
	}\href {https://doi.org/10.1126/science.1257219} {\bibfield  {journal}
		{\bibinfo  {journal} {Science}\ }\textbf {\bibinfo {volume} {346}},\ 
		\bibinfo
		{pages} {207} (\bibinfo {year} {2014})}\BibitemShut {NoStop}%
	\bibitem [{\citenamefont {Aref}\ \emph {et~al.}(2016)\citenamefont {Aref},
		\citenamefont {Delsing}, \citenamefont {Ekstr{\"{o}}m}, \citenamefont
		{Kockum}, \citenamefont {Gustafsson}, \citenamefont {Johansson},
		\citenamefont {Leek}, \citenamefont {Magnusson},\ and\ \citenamefont
		{Manenti}}]{Aref2016}%
	\BibitemOpen
	\bibfield  {author} {\bibinfo {author} {\bibfnamefont {T.}~\bibnamefont
			{Aref}}, \bibinfo {author} {\bibfnamefont {P.}~\bibnamefont 
			{Delsing}},
		\bibinfo {author} {\bibfnamefont {M.~K.}\ \bibnamefont 
		{Ekstr{\"{o}}m}},
		\bibinfo {author} {\bibfnamefont {A.~F.}\ \bibnamefont {Kockum}}, 
		\bibinfo
		{author} {\bibfnamefont {M.~V.}\ \bibnamefont {Gustafsson}}, \bibinfo
		{author} {\bibfnamefont {G.}~\bibnamefont {Johansson}}, \bibinfo 
		{author}
		{\bibfnamefont {P.~J.}\ \bibnamefont {Leek}}, \bibinfo {author}
		{\bibfnamefont {E.}~\bibnamefont {Magnusson}},\ and\ \bibinfo {author}
		{\bibfnamefont {R.}~\bibnamefont {Manenti}},\ }\bibfield  {title} 
		{\bibinfo
		{title} {{Quantum Acoustics with Surface Acoustic Waves}},\ }in\ \href
	{https://doi.org/10.1007/978-3-319-24091-6_9} {\emph {\bibinfo {booktitle}
			{Superconducting Devices in Quantum Optics}}},\ \bibinfo {editor} 
			{edited by\
		\bibinfo {editor} {\bibfnamefont {R.~H.}\ \bibnamefont {Hadfield}}\ 
		and\
		\bibinfo {editor} {\bibfnamefont {G.}~\bibnamefont {Johansson}}}\ 
		(\bibinfo
	{publisher} {Springer},\ \bibinfo {year} {2016})\BibitemShut {NoStop}%
	\bibitem [{\citenamefont {Guo}\ \emph {et~al.}(2017)\citenamefont {Guo},
		\citenamefont {Grimsmo}, \citenamefont {Kockum}, \citenamefont 
		{Pletyukhov},\
		and\ \citenamefont {Johansson}}]{Guo2017}%
	\BibitemOpen
	\bibfield  {author} {\bibinfo {author} {\bibfnamefont {L.}~\bibnamefont
			{Guo}}, \bibinfo {author} {\bibfnamefont {A.}~\bibnamefont 
			{Grimsmo}},
		\bibinfo {author} {\bibfnamefont {A.~F.}\ \bibnamefont {Kockum}}, 
		\bibinfo
		{author} {\bibfnamefont {M.}~\bibnamefont {Pletyukhov}},\ and\ 
		\bibinfo
		{author} {\bibfnamefont {G.}~\bibnamefont {Johansson}},\ }\bibfield  
		{title}
	{\bibinfo {title} {{Giant acoustic atom: A single quantum system with a
				deterministic time delay}},\ }\href
	{https://doi.org/10.1103/PhysRevA.95.053821} {\bibfield  {journal} 
	{\bibinfo
			{journal} {Phys. Rev. A}\ }\textbf {\bibinfo {volume} {95}},\ 
			\bibinfo
		{pages} {053821} (\bibinfo {year} {2017})}\BibitemShut {NoStop}%
	\bibitem [{\citenamefont {Andersson}\ \emph {et~al.}(2019)\citenamefont
		{Andersson}, \citenamefont {Suri}, \citenamefont {Guo}, \citenamefont
		{Aref},\ and\ \citenamefont {Delsing}}]{Andersson2019}%
	\BibitemOpen
	\bibfield  {author} {\bibinfo {author} {\bibfnamefont {G.}~\bibnamefont
			{Andersson}}, \bibinfo {author} {\bibfnamefont {B.}~\bibnamefont 
			{Suri}},
		\bibinfo {author} {\bibfnamefont {L.}~\bibnamefont {Guo}}, \bibinfo 
		{author}
		{\bibfnamefont {T.}~\bibnamefont {Aref}},\ and\ \bibinfo {author}
		{\bibfnamefont {P.}~\bibnamefont {Delsing}},\ }\bibfield  {title} 
		{\bibinfo
		{title} {Non-exponential decay of a giant artificial atom},\ }\href
	{https://doi.org/10.1038/s41567-019-0605-6} {\bibfield  {journal} 
	{\bibinfo
			{journal} {Nature Physics}\ }\textbf {\bibinfo {volume} {15}},\ 
			\bibinfo
		{pages} {1123} (\bibinfo {year} {2019})}\BibitemShut {NoStop}%
	\bibitem [{\citenamefont {Vadiraj}\ \emph {et~al.}(2021)\citenamefont
		{Vadiraj}, \citenamefont {Ask}, \citenamefont {McConkey}, 
		\citenamefont
		{Nsanzineza}, \citenamefont {Chang}, \citenamefont {Kockum},\ and\
		\citenamefont {Wilson}}]{Vadiraj2021}%
	\BibitemOpen
	\bibfield  {author} {\bibinfo {author} {\bibfnamefont {A.~M.}\ 
	\bibnamefont
			{Vadiraj}}, \bibinfo {author} {\bibfnamefont {A.}~\bibnamefont 
			{Ask}},
		\bibinfo {author} {\bibfnamefont {T.~G.}\ \bibnamefont {McConkey}}, 
		\bibinfo
		{author} {\bibfnamefont {I.}~\bibnamefont {Nsanzineza}}, \bibinfo 
		{author}
		{\bibfnamefont {C.~W.~S.}\ \bibnamefont {Chang}}, \bibinfo {author}
		{\bibfnamefont {A.~F.}\ \bibnamefont {Kockum}},\ and\ \bibinfo 
		{author}
		{\bibfnamefont {C.~M.}\ \bibnamefont {Wilson}},\ }\bibfield  {title}
	{\bibinfo {title} {Engineering the level structure of a giant artificial 
	atom
			in waveguide quantum electrodynamics},\ }\href
	{https://doi.org/10.1103/PhysRevA.103.023710} {\bibfield  {journal} 
	{\bibinfo
			{journal} {Phys. Rev. A}\ }\textbf {\bibinfo {volume} {103}},\ 
			\bibinfo
		{pages} {023710} (\bibinfo {year} {2021})}\BibitemShut {NoStop}%
	\bibitem [{\citenamefont {Wang}\ and\ \citenamefont 
	{Li}(2022)}]{Wang2022}%
	\BibitemOpen
	\bibfield  {author} {\bibinfo {author} {\bibfnamefont {X.}~\bibnamefont
			{Wang}}\ and\ \bibinfo {author} {\bibfnamefont {H.-R.}\ 
			\bibnamefont {Li}},\
	}\bibfield  {title} {\bibinfo {title} {Chiral quantum network with giant
			atoms},\ }\href {https://doi.org/10.1088/2058-9565/ac6a04} 
			{\bibfield
		{journal} {\bibinfo  {journal} {Quantum Sci. Technol.}\ }\textbf 
		{\bibinfo
			{volume} {7}},\ \bibinfo {pages} {035007} (\bibinfo {year}
		{2022})}\BibitemShut {NoStop}%
	\bibitem [{\citenamefont {Wang}\ \emph {et~al.}(2022)\citenamefont {Wang},
		\citenamefont {Wang}, \citenamefont {Yao}, \citenamefont {Shen},
		\citenamefont {Wu}, \citenamefont {Qian}, \citenamefont {Li}, 
		\citenamefont
		{Zhu},\ and\ \citenamefont {You}}]{WangZiQi2022}%
	\BibitemOpen
	\bibfield  {author} {\bibinfo {author} {\bibfnamefont {Z.-Q.}\ 
	\bibnamefont
			{Wang}}, \bibinfo {author} {\bibfnamefont {Y.-P.}\ \bibnamefont 
			{Wang}},
		\bibinfo {author} {\bibfnamefont {J.}~\bibnamefont {Yao}}, \bibinfo 
		{author}
		{\bibfnamefont {R.-C.}\ \bibnamefont {Shen}}, \bibinfo {author}
		{\bibfnamefont {W.-J.}\ \bibnamefont {Wu}}, \bibinfo {author} 
		{\bibfnamefont
			{J.}~\bibnamefont {Qian}}, \bibinfo {author} {\bibfnamefont 
			{J.}~\bibnamefont
			{Li}}, \bibinfo {author} {\bibfnamefont {S.-Y.}\ \bibnamefont 
			{Zhu}},\ and\
		\bibinfo {author} {\bibfnamefont {J.~Q.}\ \bibnamefont {You}},\ 
		}\bibfield
	{title} {\bibinfo {title} {Giant spin ensembles in waveguide magnonics},\
	}\href {https://doi.org/10.1038/s41467-022-35174-9} {\bibfield  {journal}
		{\bibinfo  {journal} {Nature Communications}\ }\textbf {\bibinfo 
		{volume}
			{13}},\ \bibinfo {pages} {7580} (\bibinfo {year} 
			{2022})}\BibitemShut
	{NoStop}%
	\bibitem [{\citenamefont {Kockum}\ \emph {et~al.}(2018)\citenamefont 
	{Kockum},
		\citenamefont {Johansson},\ and\ \citenamefont {Nori}}]{Anton2018}%
	\BibitemOpen
	\bibfield  {author} {\bibinfo {author} {\bibfnamefont {A.~F.}\ 
	\bibnamefont
			{Kockum}}, \bibinfo {author} {\bibfnamefont {G.}~\bibnamefont 
			{Johansson}},\
		and\ \bibinfo {author} {\bibfnamefont {F.}~\bibnamefont {Nori}},\ 
		}\bibfield
	{title} {\bibinfo {title} {{Decoherence-Free Interaction between Giant 
	Atoms
				in Waveguide Quantum Electrodynamics}},\ }\href
	{https://doi.org/10.1103/PhysRevLett.120.140404} {\bibfield  {journal}
		{\bibinfo  {journal} {Phys. Rev. Lett.}\ }\textbf {\bibinfo {volume} 
		{120}},\
		\bibinfo {pages} {140404} (\bibinfo {year} {2018})}\BibitemShut 
		{NoStop}%
	\bibitem [{\citenamefont {Kannan}\ \emph {et~al.}(2020)\citenamefont 
	{Kannan},
		\citenamefont {Ruckriegel}, \citenamefont {Campbell}, \citenamefont
		{Frisk~Kockum}, \citenamefont {Braum\"{u}ller}, \citenamefont {Kim},
		\citenamefont {Kjaergaard}, \citenamefont {Krantz}, \citenamefont 
		{Melville},
		\citenamefont {Niedzielski}, \citenamefont {Veps\"{a}l\"{a}inen},
		\citenamefont {Winik}, \citenamefont {Yoder}, \citenamefont {Nori},
		\citenamefont {Orlando}, \citenamefont {Gustavsson},\ and\ 
		\citenamefont
		{Oliver}}]{Kannan2020}%
	\BibitemOpen
	\bibfield  {author} {\bibinfo {author} {\bibfnamefont {B.}~\bibnamefont
			{Kannan}}, \bibinfo {author} {\bibfnamefont {M.~J.}\ \bibnamefont
			{Ruckriegel}}, \bibinfo {author} {\bibfnamefont {D.~L.}\ 
			\bibnamefont
			{Campbell}}, \bibinfo {author} {\bibfnamefont {A.}~\bibnamefont
			{Frisk~Kockum}}, \bibinfo {author} {\bibfnamefont 
			{J.}~\bibnamefont
			{Braum\"{u}ller}}, \bibinfo {author} {\bibfnamefont {D.~K.}\ 
			\bibnamefont
			{Kim}}, \bibinfo {author} {\bibfnamefont {M.}~\bibnamefont 
			{Kjaergaard}},
		\bibinfo {author} {\bibfnamefont {P.}~\bibnamefont {Krantz}}, \bibinfo
		{author} {\bibfnamefont {A.}~\bibnamefont {Melville}}, \bibinfo 
		{author}
		{\bibfnamefont {B.~M.}\ \bibnamefont {Niedzielski}}, \bibinfo {author}
		{\bibfnamefont {A.}~\bibnamefont {Veps\"{a}l\"{a}inen}}, \bibinfo 
		{author}
		{\bibfnamefont {R.}~\bibnamefont {Winik}}, \bibinfo {author} 
		{\bibfnamefont
			{J.~L.}\ \bibnamefont {Yoder}}, \bibinfo {author} {\bibfnamefont
			{F.}~\bibnamefont {Nori}}, \bibinfo {author} {\bibfnamefont 
			{T.~P.}\
			\bibnamefont {Orlando}}, \bibinfo {author} {\bibfnamefont 
			{S.}~\bibnamefont
			{Gustavsson}},\ and\ \bibinfo {author} {\bibfnamefont {W.~D.}\ 
			\bibnamefont
			{Oliver}},\ }\bibfield  {title} {\bibinfo {title} {Waveguide 
			quantum
			electrodynamics with superconducting artificial giant atoms},\ 
			}\href
	{https://doi.org/10.1038/s41586-020-2529-9} {\bibfield  {journal} 
	{\bibinfo
			{journal} {Nature}\ }\textbf {\bibinfo {volume} {583}},\ \bibinfo 
			{pages}
		{775} (\bibinfo {year} {2020})}\BibitemShut {NoStop}%
	\bibitem [{\citenamefont {Carollo}\ \emph {et~al.}(2020)\citenamefont
		{Carollo}, \citenamefont {Cilluffo},\ and\ \citenamefont
		{Ciccarello}}]{Carollo2020}%
	\BibitemOpen
	\bibfield  {author} {\bibinfo {author} {\bibfnamefont {A.}~\bibnamefont
			{Carollo}}, \bibinfo {author} {\bibfnamefont {D.}~\bibnamefont 
			{Cilluffo}},\
		and\ \bibinfo {author} {\bibfnamefont {F.}~\bibnamefont 
		{Ciccarello}},\
	}\bibfield  {title} {\bibinfo {title} {Mechanism of decoherence-free 
	coupling
			between giant atoms},\ }\href
	{https://doi.org/10.1103/PhysRevResearch.2.043184} {\bibfield  {journal}
		{\bibinfo  {journal} {Phys. Rev. Res.}\ }\textbf {\bibinfo {volume} 
		{2}},\
		\bibinfo {pages} {043184} (\bibinfo {year} {2020})}\BibitemShut 
		{NoStop}%
	\bibitem [{\citenamefont {Soro}\ \emph {et~al.}(2023)\citenamefont {Soro},
		\citenamefont {Mu\~noz},\ and\ \citenamefont {Kockum}}]{Soro2023}%
	\BibitemOpen
	\bibfield  {author} {\bibinfo {author} {\bibfnamefont {A.}~\bibnamefont
			{Soro}}, \bibinfo {author} {\bibfnamefont {C.~S.}\ \bibnamefont 
			{Mu\~noz}},\
		and\ \bibinfo {author} {\bibfnamefont {A.~F.}\ \bibnamefont 
		{Kockum}},\
	}\bibfield  {title} {\bibinfo {title} {Interaction between giant atoms in 
	a
			one-dimensional structured environment},\ }\href
	{https://doi.org/10.1103/PhysRevA.107.013710} {\bibfield  {journal} 
	{\bibinfo
			{journal} {Phys. Rev. A}\ }\textbf {\bibinfo {volume} {107}},\ 
			\bibinfo
		{pages} {013710} (\bibinfo {year} {2023})}\BibitemShut {NoStop}%
	\bibitem [{\citenamefont {Guo}\ \emph {et~al.}(2020)\citenamefont {Guo},
		\citenamefont {Kockum}, \citenamefont {Marquardt},\ and\ \citenamefont
		{Johansson}}]{Guo2020}%
	\BibitemOpen
	\bibfield  {author} {\bibinfo {author} {\bibfnamefont {L.}~\bibnamefont
			{Guo}}, \bibinfo {author} {\bibfnamefont {A.~F.}\ \bibnamefont 
			{Kockum}},
		\bibinfo {author} {\bibfnamefont {F.}~\bibnamefont {Marquardt}},\ and\
		\bibinfo {author} {\bibfnamefont {G.}~\bibnamefont {Johansson}},\ 
		}\bibfield
	{title} {\bibinfo {title} {Oscillating bound states for a giant atom},\
	}\href {https://doi.org/10.1103/PhysRevResearch.2.043014} {\bibfield
		{journal} {\bibinfo  {journal} {Phys. Rev. Res.}\ }\textbf {\bibinfo 
		{volume}
			{2}},\ \bibinfo {pages} {043014} (\bibinfo {year} 
			{2020})}\BibitemShut
	{NoStop}%
	\bibitem [{\citenamefont {Noachtar}\ \emph {et~al.}(2022)\citenamefont
		{Noachtar}, \citenamefont {Kn{\"{o}}rzer},\ and\ \citenamefont
		{Jonsson}}]{Noachtar2022}%
	\BibitemOpen
	\bibfield  {author} {\bibinfo {author} {\bibfnamefont {D.~D.}\ 
	\bibnamefont
			{Noachtar}}, \bibinfo {author} {\bibfnamefont {J.}~\bibnamefont
			{Kn{\"{o}}rzer}},\ and\ \bibinfo {author} {\bibfnamefont {R.~H.}\
			\bibnamefont {Jonsson}},\ }\bibfield  {title} {\bibinfo {title}
		{{Nonperturbative treatment of giant atoms using chain 
		transformations}},\
	}\href {https://doi.org/10.1103/PhysRevA.106.013702} {\bibfield  {journal}
		{\bibinfo  {journal} {Phys. Rev. A}\ }\textbf {\bibinfo {volume} 
		{106}},\
		\bibinfo {pages} {013702} (\bibinfo {year} {2022})}\BibitemShut 
		{NoStop}%
	\bibitem [{\citenamefont {Lim}\ \emph {et~al.}(2023)\citenamefont {Lim},
		\citenamefont {Mok},\ and\ \citenamefont {Kwek}}]{Lim2023}%
	\BibitemOpen
	\bibfield  {author} {\bibinfo {author} {\bibfnamefont {K.~H.}\ 
	\bibnamefont
			{Lim}}, \bibinfo {author} {\bibfnamefont {W.~K.}\ \bibnamefont 
			{Mok}},\ and\
		\bibinfo {author} {\bibfnamefont {L.~C.}\ \bibnamefont {Kwek}},\ 
		}\bibfield
	{title} {\bibinfo {title} {{Oscillating bound states in non-Markovian
				photonic lattices}},\ }\href 
				{https://doi.org/10.1103/PhysRevA.107.023716}
	{\bibfield  {journal} {\bibinfo  {journal} {Phys. Rev. A}\ }\textbf 
	{\bibinfo
			{volume} {107}},\ \bibinfo {pages} {023716} (\bibinfo {year}
		{2023})}\BibitemShut {NoStop}%
	\bibitem [{\citenamefont {Zhao}\ and\ \citenamefont 
	{Wang}(2020)}]{Wei2020}%
	\BibitemOpen
	\bibfield  {author} {\bibinfo {author} {\bibfnamefont {W.}~\bibnamefont
			{Zhao}}\ and\ \bibinfo {author} {\bibfnamefont {Z.}~\bibnamefont 
			{Wang}},\
	}\bibfield  {title} {\bibinfo {title} {Single-photon scattering and bound
			states in an atom-waveguide system with two or multiple coupling 
			points},\
	}\href {https://doi.org/10.1103/PhysRevA.101.053855} {\bibfield  {journal}
		{\bibinfo  {journal} {Phys. Rev. A}\ }\textbf {\bibinfo {volume} 
		{101}},\
		\bibinfo {pages} {053855} (\bibinfo {year} {2020})}\BibitemShut 
		{NoStop}%
	\bibitem [{\citenamefont {Gonz\'alez-Tudela}\ \emph 
	{et~al.}(2019)\citenamefont
		{Gonz\'alez-Tudela}, \citenamefont {Mu\~noz},\ and\ \citenamefont
		{Cirac}}]{Gonzalez2019}%
	\BibitemOpen
	\bibfield  {author} {\bibinfo {author} {\bibfnamefont {A.}~\bibnamefont
			{Gonz\'alez-Tudela}}, \bibinfo {author} {\bibfnamefont {C.~S.}\ 
			\bibnamefont
			{Mu\~noz}},\ and\ \bibinfo {author} {\bibfnamefont {J.~I.}\ 
			\bibnamefont
			{Cirac}},\ }\bibfield  {title} {\bibinfo {title} {{Engineering 
			and Harnessing
				Giant Atoms in High-Dimensional Baths: A Proposal for 
				Implementation with
				Cold Atoms}},\ }\href 
				{https://doi.org/10.1103/PhysRevLett.122.203603}
	{\bibfield  {journal} {\bibinfo  {journal} {Phys. Rev. Lett.}\ }\textbf
		{\bibinfo {volume} {122}},\ \bibinfo {pages} {203603} (\bibinfo {year}
		{2019})}\BibitemShut {NoStop}%
	\bibitem [{\citenamefont {Wang}\ \emph {et~al.}(2021)\citenamefont {Wang},
		\citenamefont {Liu}, \citenamefont {Kockum}, \citenamefont {Li},\ and\
		\citenamefont {Nori}}]{Wang2021}%
	\BibitemOpen
	\bibfield  {author} {\bibinfo {author} {\bibfnamefont {X.}~\bibnamefont
			{Wang}}, \bibinfo {author} {\bibfnamefont {T.}~\bibnamefont 
			{Liu}}, \bibinfo
		{author} {\bibfnamefont {A.~F.}\ \bibnamefont {Kockum}}, \bibinfo 
		{author}
		{\bibfnamefont {H.-R.}\ \bibnamefont {Li}},\ and\ \bibinfo {author}
		{\bibfnamefont {F.}~\bibnamefont {Nori}},\ }\bibfield  {title} 
		{\bibinfo
		{title} {{Tunable Chiral Bound States with Giant Atoms}},\ }\href
	{https://doi.org/10.1103/PhysRevLett.126.043602} {\bibfield  {journal}
		{\bibinfo  {journal} {Phys. Rev. Lett.}\ }\textbf {\bibinfo {volume} 
		{126}},\
		\bibinfo {pages} {043602} (\bibinfo {year} {2021})}\BibitemShut 
		{NoStop}%
	\bibitem [{\citenamefont {Joshi}\ \emph {et~al.}(2023)\citenamefont 
	{Joshi},
		\citenamefont {Yang},\ and\ \citenamefont 
		{Mirhosseini}}]{Chaitali2023}%
	\BibitemOpen
	\bibfield  {author} {\bibinfo {author} {\bibfnamefont {C.}~\bibnamefont
			{Joshi}}, \bibinfo {author} {\bibfnamefont {F.}~\bibnamefont 
			{Yang}},\ and\
		\bibinfo {author} {\bibfnamefont {M.}~\bibnamefont {Mirhosseini}},\
	}\bibfield  {title} {\bibinfo {title} {{Resonance Fluorescence of a Chiral
				Artificial Atom}},\ }\href 
				{https://doi.org/10.1103/PhysRevX.13.021039}
	{\bibfield  {journal} {\bibinfo  {journal} {Phys. Rev. X}\ }\textbf 
	{\bibinfo
			{volume} {13}},\ \bibinfo {pages} {021039} (\bibinfo {year}
		{2023})}\BibitemShut {NoStop}%
	\bibitem [{\citenamefont {Scully}\ and\ \citenamefont
		{Zubairy}(1997)}]{Scully1997}%
	\BibitemOpen
	\bibfield  {author} {\bibinfo {author} {\bibfnamefont {M.~O.}\ 
	\bibnamefont
			{Scully}}\ and\ \bibinfo {author} {\bibfnamefont {M.~S.}\ 
			\bibnamefont
			{Zubairy}},\ }\href@noop {} {\emph {\bibinfo {title} {Quantum 
			optics}}}\
	(\bibinfo  {publisher} {Cambridge University Press},\ \bibinfo {year}
	{1997})\BibitemShut {NoStop}%
	\bibitem [{\citenamefont {Ramos}\ \emph {et~al.}(2016)\citenamefont 
	{Ramos},
		\citenamefont {Vermersch}, \citenamefont {Hauke}, \citenamefont 
		{Pichler},\
		and\ \citenamefont {Zoller}}]{Ramos2016}%
	\BibitemOpen
	\bibfield  {author} {\bibinfo {author} {\bibfnamefont {T.}~\bibnamefont
			{Ramos}}, \bibinfo {author} {\bibfnamefont {B.}~\bibnamefont 
			{Vermersch}},
		\bibinfo {author} {\bibfnamefont {P.}~\bibnamefont {Hauke}}, \bibinfo
		{author} {\bibfnamefont {H.}~\bibnamefont {Pichler}},\ and\ \bibinfo 
		{author}
		{\bibfnamefont {P.}~\bibnamefont {Zoller}},\ }\bibfield  {title} 
		{\bibinfo
		{title} {{Non-Markovian dynamics in chiral quantum networks with 
		spins and
				photons}},\ }\href 
				{https://doi.org/10.1103/PhysRevA.93.062104} {\bibfield
		{journal} {\bibinfo  {journal} {Phys. Rev. A}\ }\textbf {\bibinfo 
		{volume}
			{93}},\ \bibinfo {pages} {062104} (\bibinfo {year} 
			{2016})}\BibitemShut
	{NoStop}%
	\bibitem [{\citenamefont {Wang}\ \emph {et~al.}(2024)\citenamefont {Wang},
		\citenamefont {Zhu}, \citenamefont {Liu},\ and\ \citenamefont
		{Nori}}]{Wang2024}%
	\BibitemOpen
	\bibfield  {author} {\bibinfo {author} {\bibfnamefont {X.}~\bibnamefont
			{Wang}}, \bibinfo {author} {\bibfnamefont {H.-B.}\ \bibnamefont 
			{Zhu}},
		\bibinfo {author} {\bibfnamefont {T.}~\bibnamefont {Liu}},\ and\ 
		\bibinfo
		{author} {\bibfnamefont {F.}~\bibnamefont {Nori}},\ }\bibfield  
		{title}
	{\bibinfo {title} {Realizing quantum optics in structured environments 
	with
			giant atoms},\ }\href 
			{https://doi.org/10.1103/PhysRevResearch.6.013279}
	{\bibfield  {journal} {\bibinfo  {journal} {Phys. Rev. Res.}\ }\textbf
		{\bibinfo {volume} {6}},\ \bibinfo {pages} {013279} (\bibinfo {year}
		{2024})}\BibitemShut {NoStop}%
	\bibitem [{\citenamefont {Gao}\ \emph {et~al.}(2024)\citenamefont {Gao},
		\citenamefont {Li}, \citenamefont {Wu}, \citenamefont {Liu},\ and\
		\citenamefont {Wang}}]{Gao2024}%
	\BibitemOpen
	\bibfield  {author} {\bibinfo {author} {\bibfnamefont {Z.-M.}\ 
	\bibnamefont
			{Gao}}, \bibinfo {author} {\bibfnamefont {J.-Q.}\ \bibnamefont 
			{Li}},
		\bibinfo {author} {\bibfnamefont {Y.-H.}\ \bibnamefont {Wu}}, \bibinfo
		{author} {\bibfnamefont {W.-X.}\ \bibnamefont {Liu}},\ and\ \bibinfo 
		{author}
		{\bibfnamefont {X.}~\bibnamefont {Wang}},\ }\bibfield  {title} 
		{\bibinfo
		{title} {Harnessing spontaneous emission of correlated photon pairs 
		from
			ladder-type giant atoms},\ }\href
	{https://doi.org/10.1103/PhysRevA.110.053706} {\bibfield  {journal} 
	{\bibinfo
			{journal} {Phys. Rev. A}\ }\textbf {\bibinfo {volume} {110}},\ 
			\bibinfo
		{pages} {053706} (\bibinfo {year} {2024})}\BibitemShut {NoStop}%
	\bibitem [{\citenamefont {Wang}\ \emph {et~al.}(2015)\citenamefont {Wang},
		\citenamefont {Zhu}, \citenamefont {Evers},\ and\ \citenamefont
		{Scully}}]{WangDawei2015}%
	\BibitemOpen
	\bibfield  {author} {\bibinfo {author} {\bibfnamefont {D.-W.}\ 
	\bibnamefont
			{Wang}}, \bibinfo {author} {\bibfnamefont {S.-Y.}\ \bibnamefont 
			{Zhu}},
		\bibinfo {author} {\bibfnamefont {J.}~\bibnamefont {Evers}},\ and\ 
		\bibinfo
		{author} {\bibfnamefont {M.~O.}\ \bibnamefont {Scully}},\ }\bibfield  
		{title}
	{\bibinfo {title} {High-frequency light reflector via low-frequency light
			control},\ }\href {https://doi.org/10.1103/PhysRevA.91.011801} 
			{\bibfield
		{journal} {\bibinfo  {journal} {Phys. Rev. A}\ }\textbf {\bibinfo 
		{volume}
			{91}},\ \bibinfo {pages} {011801} (\bibinfo {year} 
			{2015})}\BibitemShut
	{NoStop}%
	\bibitem [{\citenamefont {Zhong}\ \emph {et~al.}(2021)\citenamefont 
	{Zhong},
		\citenamefont {Chang}, \citenamefont {Bienfait}, \citenamefont 
		{Dumur},
		\citenamefont {Chou}, \citenamefont {Conner}, \citenamefont {Grebel},
		\citenamefont {Povey}, \citenamefont {Yan}, \citenamefont 
		{Schuster},\ and\
		\citenamefont {Cleland}}]{Zhong2021}%
	\BibitemOpen
	\bibfield  {author} {\bibinfo {author} {\bibfnamefont {Y.}~\bibnamefont
			{Zhong}}, \bibinfo {author} {\bibfnamefont {H.-S.}\ \bibnamefont 
			{Chang}},
		\bibinfo {author} {\bibfnamefont {A.}~\bibnamefont {Bienfait}}, 
		\bibinfo
		{author} {\bibfnamefont {{\'E}.}~\bibnamefont {Dumur}}, \bibinfo 
		{author}
		{\bibfnamefont {M.-H.}\ \bibnamefont {Chou}}, \bibinfo {author}
		{\bibfnamefont {C.~R.}\ \bibnamefont {Conner}}, \bibinfo {author}
		{\bibfnamefont {J.}~\bibnamefont {Grebel}}, \bibinfo {author} 
		{\bibfnamefont
			{R.~G.}\ \bibnamefont {Povey}}, \bibinfo {author} {\bibfnamefont
			{H.}~\bibnamefont {Yan}}, \bibinfo {author} {\bibfnamefont 
			{D.~I.}\
			\bibnamefont {Schuster}},\ and\ \bibinfo {author} {\bibfnamefont 
			{A.~N.}\
			\bibnamefont {Cleland}},\ }\bibfield  {title} {\bibinfo {title}
		{Deterministic multi-qubit entanglement in a quantum network},\ }\href
	{https://doi.org/10.1038/s41586-021-03288-7} {\bibfield  {journal} 
	{\bibinfo
			{journal} {Nature}\ }\textbf {\bibinfo {volume} {590}},\ \bibinfo 
			{pages}
		{571} (\bibinfo {year} {2021})}\BibitemShut {NoStop}%
	\bibitem [{\citenamefont {Strand}\ \emph {et~al.}(2015)\citenamefont 
	{Strand},
		\citenamefont {Eckstein},\ and\ \citenamefont {Werner}}]{Strand2015}%
	\BibitemOpen
	\bibfield  {author} {\bibinfo {author} {\bibfnamefont {H.~U.~R.}\
			\bibnamefont {Strand}}, \bibinfo {author} {\bibfnamefont 
			{M.}~\bibnamefont
			{Eckstein}},\ and\ \bibinfo {author} {\bibfnamefont 
			{P.}~\bibnamefont
			{Werner}},\ }\bibfield  {title} {\bibinfo {title} {Beyond the 
			hubbard bands
			in strongly correlated lattice bosons},\ }\href
	{https://doi.org/10.1103/PhysRevA.92.063602} {\bibfield  {journal} 
	{\bibinfo
			{journal} {Phys. Rev. A}\ }\textbf {\bibinfo {volume} {92}},\ 
			\bibinfo
		{pages} {063602} (\bibinfo {year} {2015})}\BibitemShut {NoStop}%
	\bibitem [{\citenamefont {Sajna}(2016)}]{Sajna2016}%
	\BibitemOpen
	\bibfield  {author} {\bibinfo {author} {\bibfnamefont {A.~S.}\ 
	\bibnamefont
			{Sajna}},\ }\bibfield  {title} {\bibinfo {title} {Effects of 
			higher-order
			energy bands and temperature on the bosonic mott insulator in a 
			periodically
			modulated lattice},\ }\href 
			{https://doi.org/10.1103/PhysRevA.94.043612}
	{\bibfield  {journal} {\bibinfo  {journal} {Phys. Rev. A}\ }\textbf 
	{\bibinfo
			{volume} {94}},\ \bibinfo {pages} {043612} (\bibinfo {year}
		{2016})}\BibitemShut {NoStop}%
	\bibitem [{\citenamefont {Chang}\ \emph {et~al.}(2020)\citenamefont 
	{Chang},
		\citenamefont {Sab\'{\i}n}, \citenamefont {Forn-D\'{\i}az}, 
		\citenamefont
		{Quijandr\'{\i}a}, \citenamefont {Vadiraj}, \citenamefont 
		{Nsanzineza},
		\citenamefont {Johansson},\ and\ \citenamefont {Wilson}}]{Chang2020x}%
	\BibitemOpen
	\bibfield  {author} {\bibinfo {author} {\bibfnamefont {C.~W.~S.}\
			\bibnamefont {Chang}}, \bibinfo {author} {\bibfnamefont 
			{C.}~\bibnamefont
			{Sab\'{\i}n}}, \bibinfo {author} {\bibfnamefont {P.}~\bibnamefont
			{Forn-D\'{\i}az}}, \bibinfo {author} {\bibfnamefont 
			{F.}~\bibnamefont
			{Quijandr\'{\i}a}}, \bibinfo {author} {\bibfnamefont {A.~M.}\ 
			\bibnamefont
			{Vadiraj}}, \bibinfo {author} {\bibfnamefont {I.}~\bibnamefont 
			{Nsanzineza}},
		\bibinfo {author} {\bibfnamefont {G.}~\bibnamefont {Johansson}},\ and\
		\bibinfo {author} {\bibfnamefont {C.~M.}\ \bibnamefont {Wilson}},\ 
		}\bibfield
	{title} {\bibinfo {title} {Observation of three-photon spontaneous
			parametric down-conversion in a superconducting parametric 
			cavity},\ }\href
	{https://doi.org/10.1103/PhysRevX.10.011011} {\bibfield  {journal} 
	{\bibinfo
			{journal} {Phys. Rev. X}\ }\textbf {\bibinfo {volume} {10}},\ 
			\bibinfo
		{pages} {011011} (\bibinfo {year} {2020})}\BibitemShut {NoStop}%
	\bibitem [{\citenamefont {Gu}\ \emph {et~al.}(2017)\citenamefont {Gu},
		\citenamefont {Kockum}, \citenamefont {Miranowicz}, \citenamefont 
		{Liu},\
		and\ \citenamefont {Nori}}]{Gu2017}%
	\BibitemOpen
	\bibfield  {author} {\bibinfo {author} {\bibfnamefont {X.}~\bibnamefont
			{Gu}}, \bibinfo {author} {\bibfnamefont {A.~F.}\ \bibnamefont 
			{Kockum}},
		\bibinfo {author} {\bibfnamefont {A.}~\bibnamefont {Miranowicz}}, 
		\bibinfo
		{author} {\bibfnamefont {Y.-X.}\ \bibnamefont {Liu}},\ and\ \bibinfo 
		{author}
		{\bibfnamefont {F.}~\bibnamefont {Nori}},\ }\bibfield  {title} 
		{\bibinfo
		{title} {Microwave photonics with superconducting quantum circuits},\ 
		}\href
	{https://doi.org/10.1016/j.physrep.2017.10.002} {\bibfield  {journal}
		{\bibinfo  {journal} {Phys. Rep.}\ }\textbf {\bibinfo {volume} 
		{718-719}},\
		\bibinfo {pages} {1} (\bibinfo {year} {2017})}\BibitemShut {NoStop}%
	\bibitem [{\citenamefont {Krantz}\ \emph {et~al.}(2019)\citenamefont 
	{Krantz},
		\citenamefont {Kjaergaard}, \citenamefont {Yan}, \citenamefont 
		{Orlando},
		\citenamefont {Gustavsson},\ and\ \citenamefont 
		{Oliver}}]{Krantz2019}%
	\BibitemOpen
	\bibfield  {author} {\bibinfo {author} {\bibfnamefont {P.}~\bibnamefont
			{Krantz}}, \bibinfo {author} {\bibfnamefont {M.}~\bibnamefont 
			{Kjaergaard}},
		\bibinfo {author} {\bibfnamefont {F.}~\bibnamefont {Yan}}, \bibinfo 
		{author}
		{\bibfnamefont {T.~P.}\ \bibnamefont {Orlando}}, \bibinfo {author}
		{\bibfnamefont {S.}~\bibnamefont {Gustavsson}},\ and\ \bibinfo 
		{author}
		{\bibfnamefont {W.~D.}\ \bibnamefont {Oliver}},\ }\bibfield  {title}
	{\bibinfo {title} {{A quantum engineer's guide to superconducting 
	qubits}},\
	}\href {https://doi.org/10.1063/1.5089550} {\bibfield  {journal} {\bibinfo
			{journal} {Appl. Phys. Rev.}\ }\textbf {\bibinfo {volume} {6}},\ 
			\bibinfo
		{pages} {021318} (\bibinfo {year} {2019})}\BibitemShut {NoStop}%
	\bibitem [{\citenamefont {Hacohen-Gourgy}\ \emph 
	{et~al.}(2015)\citenamefont
		{Hacohen-Gourgy}, \citenamefont {Ramasesh}, \citenamefont {De~Grandi},
		\citenamefont {Siddiqi},\ and\ \citenamefont {Girvin}}]{Gourgy2015}%
	\BibitemOpen
	\bibfield  {author} {\bibinfo {author} {\bibfnamefont {S.}~\bibnamefont
			{Hacohen-Gourgy}}, \bibinfo {author} {\bibfnamefont {V.~V.}\ 
			\bibnamefont
			{Ramasesh}}, \bibinfo {author} {\bibfnamefont {C.}~\bibnamefont 
			{De~Grandi}},
		\bibinfo {author} {\bibfnamefont {I.}~\bibnamefont {Siddiqi}},\ and\ 
		\bibinfo
		{author} {\bibfnamefont {S.~M.}\ \bibnamefont {Girvin}},\ }\bibfield  
		{title}
	{\bibinfo {title} {Cooling and autonomous feedback in a bose-hubbard chain
			with attractive interactions},\ }\href
	{https://doi.org/10.1103/PhysRevLett.115.240501} {\bibfield  {journal}
		{\bibinfo  {journal} {Phys. Rev. Lett.}\ }\textbf {\bibinfo {volume} 
		{115}},\
		\bibinfo {pages} {240501} (\bibinfo {year} {2015})}\BibitemShut 
		{NoStop}%
	\bibitem [{\citenamefont {Roushan}\ \emph
		{et~al.}(2017{\natexlab{a}})\citenamefont {Roushan}, \citenamefont 
		{Neill},
		\citenamefont {Tangpanitanon}, \citenamefont {Bastidas}, \citenamefont
		{Megrant}, \citenamefont {Barends}, \citenamefont {Chen}, 
		\citenamefont
		{Chen}, \citenamefont {Chiaro}, \citenamefont {Dunsworth}, 
		\citenamefont
		{Fowler}, \citenamefont {Foxen}, \citenamefont {Giustina}, 
		\citenamefont
		{Jeffrey}, \citenamefont {Kelly}, \citenamefont {Lucero}, 
		\citenamefont
		{Mutus}, \citenamefont {Neeley}, \citenamefont {Quintana}, 
		\citenamefont
		{Sank}, \citenamefont {Vainsencher}, \citenamefont {Wenner}, 
		\citenamefont
		{White}, \citenamefont {Neven}, \citenamefont {Angelakis},\ and\
		\citenamefont {Martinis}}]{Roushan2017}%
	\BibitemOpen
	\bibfield  {author} {\bibinfo {author} {\bibfnamefont {P.}~\bibnamefont
			{Roushan}}, \bibinfo {author} {\bibfnamefont {C.}~\bibnamefont 
			{Neill}},
		\bibinfo {author} {\bibfnamefont {J.}~\bibnamefont {Tangpanitanon}}, 
		\bibinfo
		{author} {\bibfnamefont {V.~M.}\ \bibnamefont {Bastidas}}, \bibinfo 
		{author}
		{\bibfnamefont {A.}~\bibnamefont {Megrant}}, \bibinfo {author} 
		{\bibfnamefont
			{R.}~\bibnamefont {Barends}}, \bibinfo {author} {\bibfnamefont
			{Y.}~\bibnamefont {Chen}}, \bibinfo {author} {\bibfnamefont 
			{Z.}~\bibnamefont
			{Chen}}, \bibinfo {author} {\bibfnamefont {B.}~\bibnamefont 
			{Chiaro}},
		\bibinfo {author} {\bibfnamefont {A.}~\bibnamefont {Dunsworth}}, 
		\bibinfo
		{author} {\bibfnamefont {A.}~\bibnamefont {Fowler}}, \bibinfo {author}
		{\bibfnamefont {B.}~\bibnamefont {Foxen}}, \bibinfo {author} 
		{\bibfnamefont
			{M.}~\bibnamefont {Giustina}}, \bibinfo {author} {\bibfnamefont
			{E.}~\bibnamefont {Jeffrey}}, \bibinfo {author} {\bibfnamefont
			{J.}~\bibnamefont {Kelly}}, \bibinfo {author} {\bibfnamefont
			{E.}~\bibnamefont {Lucero}}, \bibinfo {author} {\bibfnamefont
			{J.}~\bibnamefont {Mutus}}, \bibinfo {author} {\bibfnamefont
			{M.}~\bibnamefont {Neeley}}, \bibinfo {author} {\bibfnamefont
			{C.}~\bibnamefont {Quintana}}, \bibinfo {author} {\bibfnamefont
			{D.}~\bibnamefont {Sank}}, \bibinfo {author} {\bibfnamefont 
			{A.}~\bibnamefont
			{Vainsencher}}, \bibinfo {author} {\bibfnamefont 
			{J.}~\bibnamefont {Wenner}},
		\bibinfo {author} {\bibfnamefont {T.}~\bibnamefont {White}}, \bibinfo
		{author} {\bibfnamefont {H.}~\bibnamefont {Neven}}, \bibinfo {author}
		{\bibfnamefont {D.~G.}\ \bibnamefont {Angelakis}},\ and\ \bibinfo 
		{author}
		{\bibfnamefont {J.}~\bibnamefont {Martinis}},\ }\bibfield  {title} 
		{\bibinfo
		{title} {Spectroscopic signatures of localization with interacting 
		photons in
			superconducting qubits},\ }\href 
			{https://doi.org/10.1126/science.aao1401}
	{\bibfield  {journal} {\bibinfo  {journal} {Science}\ }\textbf {\bibinfo
			{volume} {358}},\ \bibinfo {pages} {1175} (\bibinfo {year}
		{2017}{\natexlab{a}})}\BibitemShut {NoStop}%
	\bibitem [{\citenamefont {Kim}\ \emph {et~al.}(2021)\citenamefont {Kim},
		\citenamefont {Zhang}, \citenamefont {Ferreira}, \citenamefont 
		{Banker},
		\citenamefont {Iverson}, \citenamefont {Sipahigil}, \citenamefont 
		{Bello},
		\citenamefont {Gonz\'alez-Tudela}, \citenamefont {Mirhosseini},\ and\
		\citenamefont {Painter}}]{Kim2021}%
	\BibitemOpen
	\bibfield  {author} {\bibinfo {author} {\bibfnamefont {E.}~\bibnamefont
			{Kim}}, \bibinfo {author} {\bibfnamefont {X.}~\bibnamefont 
			{Zhang}}, \bibinfo
		{author} {\bibfnamefont {V.~S.}\ \bibnamefont {Ferreira}}, \bibinfo 
		{author}
		{\bibfnamefont {J.}~\bibnamefont {Banker}}, \bibinfo {author} 
		{\bibfnamefont
			{J.~K.}\ \bibnamefont {Iverson}}, \bibinfo {author} {\bibfnamefont
			{A.}~\bibnamefont {Sipahigil}}, \bibinfo {author} {\bibfnamefont
			{M.}~\bibnamefont {Bello}}, \bibinfo {author} {\bibfnamefont
			{A.}~\bibnamefont {Gonz\'alez-Tudela}}, \bibinfo {author} 
			{\bibfnamefont
			{M.}~\bibnamefont {Mirhosseini}},\ and\ \bibinfo {author} 
			{\bibfnamefont
			{O.}~\bibnamefont {Painter}},\ }\bibfield  {title} {\bibinfo 
			{title} {Quantum
			electrodynamics in a topological waveguide},\ }\href
	{https://doi.org/10.1103/PhysRevX.11.011015} {\bibfield  {journal} 
	{\bibinfo
			{journal} {Phys. Rev. X}\ }\textbf {\bibinfo {volume} {11}},\ 
			\bibinfo
		{pages} {011015} (\bibinfo {year} {2021})}\BibitemShut {NoStop}%
	\bibitem [{\citenamefont {Roushan}\ \emph
		{et~al.}(2017{\natexlab{b}})\citenamefont {Roushan} \emph
		{et~al.}}]{Roushan2017nature}%
	\BibitemOpen
	\bibfield  {author} {\bibinfo {author} {\bibfnamefont {P.}~\bibnamefont
			{Roushan}} \emph {et~al.},\ }\bibfield  {title} {\bibinfo {title} 
			{Chiral
			ground-state currents of interacting photons in a synthetic 
			magnetic field},\
	}\href {https://doi.org/10.1038/nphys3930} {\bibfield  {journal} {\bibinfo
			{journal} {Nat. Phys.}\ }\textbf {\bibinfo {volume} {13}},\ 
			\bibinfo {pages}
		{146} (\bibinfo {year} {2017}{\natexlab{b}})}\BibitemShut {NoStop}%
	\bibitem [{\citenamefont {Krinner}\ \emph {et~al.}(2022)\citenamefont
		{Krinner}, \citenamefont {Lacroix}, \citenamefont {Remm}, 
		\citenamefont
		{Di~Paolo}, \citenamefont {Genois}, \citenamefont {Leroux}, 
		\citenamefont
		{Hellings}, \citenamefont {Lazar}, \citenamefont {Swiadek}, 
		\citenamefont
		{Herrmann}, \citenamefont {Norris}, \citenamefont {Andersen}, 
		\citenamefont
		{M\"{u}ller}, \citenamefont {Blais}, \citenamefont {Eichler},\ and\
		\citenamefont {Wallraff}}]{Krinner2022m}%
	\BibitemOpen
	\bibfield  {author} {\bibinfo {author} {\bibfnamefont {S.}~\bibnamefont
			{Krinner}}, \bibinfo {author} {\bibfnamefont {N.}~\bibnamefont 
			{Lacroix}},
		\bibinfo {author} {\bibfnamefont {A.}~\bibnamefont {Remm}}, \bibinfo 
		{author}
		{\bibfnamefont {A.}~\bibnamefont {Di~Paolo}}, \bibinfo {author}
		{\bibfnamefont {E.}~\bibnamefont {Genois}}, \bibinfo {author} 
		{\bibfnamefont
			{C.}~\bibnamefont {Leroux}}, \bibinfo {author} {\bibfnamefont
			{C.}~\bibnamefont {Hellings}}, \bibinfo {author} {\bibfnamefont
			{S.}~\bibnamefont {Lazar}}, \bibinfo {author} {\bibfnamefont
			{F.}~\bibnamefont {Swiadek}}, \bibinfo {author} {\bibfnamefont
			{J.}~\bibnamefont {Herrmann}}, \bibinfo {author} {\bibfnamefont 
			{G.~J.}\
			\bibnamefont {Norris}}, \bibinfo {author} {\bibfnamefont {C.~K.}\
			\bibnamefont {Andersen}}, \bibinfo {author} {\bibfnamefont 
			{M.}~\bibnamefont
			{M\"{u}ller}}, \bibinfo {author} {\bibfnamefont {A.}~\bibnamefont 
			{Blais}},
		\bibinfo {author} {\bibfnamefont {C.}~\bibnamefont {Eichler}},\ and\ 
		\bibinfo
		{author} {\bibfnamefont {A.}~\bibnamefont {Wallraff}},\ }\bibfield  
		{title}
	{\bibinfo {title} {Realizing repeated quantum error correction in a
			distance-three surface code},\ }\href
	{https://doi.org/10.1038/s41586-022-04566-8} {\bibfield  {journal} 
	{\bibinfo
			{journal} {Nature}\ }\textbf {\bibinfo {volume} {605}},\ \bibinfo 
			{pages}
		{669} (\bibinfo {year} {2022})}\BibitemShut {NoStop}%
	\bibitem [{\citenamefont {Place}\ \emph {et~al.}(2021)\citenamefont 
	{Place},
		\citenamefont {Rodgers}, \citenamefont {Mundada}, \citenamefont 
		{Smitham},
		\citenamefont {Fitzpatrick}, \citenamefont {Leng}, \citenamefont 
		{Premkumar},
		\citenamefont {Bryon}, \citenamefont {Vrajitoarea}, \citenamefont 
		{Sussman},
		\citenamefont {Cheng}, \citenamefont {Madhavan}, \citenamefont 
		{Babla},
		\citenamefont {Le}, \citenamefont {Gang}, \citenamefont {J\"{a}ck},
		\citenamefont {Gyenis}, \citenamefont {Yao}, \citenamefont {Cava},
		\citenamefont {de~Leon},\ and\ \citenamefont {Houck}}]{Place2021m}%
	\BibitemOpen
	\bibfield  {author} {\bibinfo {author} {\bibfnamefont {A.~P.~M.}\
			\bibnamefont {Place}}, \bibinfo {author} {\bibfnamefont 
			{L.~V.~H.}\
			\bibnamefont {Rodgers}}, \bibinfo {author} {\bibfnamefont 
			{P.}~\bibnamefont
			{Mundada}}, \bibinfo {author} {\bibfnamefont {B.~M.}\ \bibnamefont
			{Smitham}}, \bibinfo {author} {\bibfnamefont {M.}~\bibnamefont
			{Fitzpatrick}}, \bibinfo {author} {\bibfnamefont 
			{Z.}~\bibnamefont {Leng}},
		\bibinfo {author} {\bibfnamefont {A.}~\bibnamefont {Premkumar}}, 
		\bibinfo
		{author} {\bibfnamefont {J.}~\bibnamefont {Bryon}}, \bibinfo {author}
		{\bibfnamefont {A.}~\bibnamefont {Vrajitoarea}}, \bibinfo {author}
		{\bibfnamefont {S.}~\bibnamefont {Sussman}}, \bibinfo {author} 
		{\bibfnamefont
			{G.}~\bibnamefont {Cheng}}, \bibinfo {author} {\bibfnamefont
			{T.}~\bibnamefont {Madhavan}}, \bibinfo {author} {\bibfnamefont 
			{H.~K.}\
			\bibnamefont {Babla}}, \bibinfo {author} {\bibfnamefont {X.~H.}\ 
			\bibnamefont
			{Le}}, \bibinfo {author} {\bibfnamefont {Y.}~\bibnamefont 
			{Gang}}, \bibinfo
		{author} {\bibfnamefont {B.}~\bibnamefont {J\"{a}ck}}, \bibinfo 
		{author}
		{\bibfnamefont {A.}~\bibnamefont {Gyenis}}, \bibinfo {author} 
		{\bibfnamefont
			{N.}~\bibnamefont {Yao}}, \bibinfo {author} {\bibfnamefont 
			{R.~J.}\
			\bibnamefont {Cava}}, \bibinfo {author} {\bibfnamefont {N.~P.}\ 
			\bibnamefont
			{de~Leon}},\ and\ \bibinfo {author} {\bibfnamefont {A.~A.}\ 
			\bibnamefont
			{Houck}},\ }\bibfield  {title} {\bibinfo {title} {New material 
			platform for
			superconducting transmon qubits with coherence times exceeding 0.3
			milliseconds},\ }\href 
			{http://dx.doi.org/10.1038/s41467-021-22030-5}
	{\bibfield  {journal} {\bibinfo  {journal} {Nature Communications}\ 
	}\textbf
		{\bibinfo {volume} {12}},\ \bibinfo {pages} {1779} (\bibinfo {year}
		{2021})}\BibitemShut {NoStop}%
	\bibitem [{\citenamefont {Johansson}\ \emph {et~al.}(2012)\citenamefont
		{Johansson}, \citenamefont {Nation},\ and\ \citenamefont
		{Nori}}]{Johansson12qutip}%
	\BibitemOpen
	\bibfield  {author} {\bibinfo {author} {\bibfnamefont {J.~R.}\ 
	\bibnamefont
			{Johansson}}, \bibinfo {author} {\bibfnamefont {P.~D.}\ 
			\bibnamefont
			{Nation}},\ and\ \bibinfo {author} {\bibfnamefont 
			{F.}~\bibnamefont {Nori}},\
	}\bibfield  {title} {\bibinfo {title} {{QuTiP: An open-source Python
				framework for the dynamics of open quantum systems}},\ }\href
	{http://www.sciencedirect.com/science/article/pii/S0010465512000835}
	{\bibfield  {journal} {\bibinfo  {journal} {Comput. Phys. Commun.}\ 
	}\textbf
		{\bibinfo {volume} {183}},\ \bibinfo {pages} {1760} (\bibinfo {year}
		{2012})}\BibitemShut {NoStop}%
	\bibitem [{\citenamefont {Johansson}\ \emph {et~al.}(2013)\citenamefont
		{Johansson}, \citenamefont {Nation},\ and\ \citenamefont
		{Nori}}]{Johansson13qutip}%
	\BibitemOpen
	\bibfield  {author} {\bibinfo {author} {\bibfnamefont {J.~R.}\ 
	\bibnamefont
			{Johansson}}, \bibinfo {author} {\bibfnamefont {P.~D.}\ 
			\bibnamefont
			{Nation}},\ and\ \bibinfo {author} {\bibfnamefont 
			{F.}~\bibnamefont {Nori}},\
	}\bibfield  {title} {\bibinfo {title} {{QuTiP 2: A Python framework for 
	the
				dynamics of open quantum systems}},\ }\href
	{http://www.sciencedirect.com/science/article/pii/S0010465512003955}
	{\bibfield  {journal} {\bibinfo  {journal} {Comput. Phys. Commun.}\ 
	}\textbf
		{\bibinfo {volume} {184}},\ \bibinfo {pages} {1234} (\bibinfo {year}
		{2013})}\BibitemShut {NoStop}%
\end{thebibliography}
%apsrev4-2.bst 2019-01-14 (MD) hand-edited version of apsrev4-1.bst
%Control: key (0)
%Control: author (8) initials jnrlst
%Control: editor formatted (1) identically to author
%Control: production of article title (0) allowed
%Control: page (0) single
%Control: year (1) truncated
%Control: production of eprint (0) enabled
%

%%%%%%%%%%%%%%%%%%%%%%%%%%%%%%%%%%%%%%%%
\vspace{.3cm}
\noindent{\large \textbf{Acknowledgements} 
\\
\noindent
X.W.~is supported by the National Natural Science 
Foundation of China (NSFC) (Grant No.~12174303). Z.H.W.~acknowledges the 
support from National Natural Science Foundation of China (Grant 
No.~12375010). T.L. acknowledges the support from Guangdong Provincial Quantum Science Strategic Initiative (Grant No.~GDZX2505004),   National Natural
Science Foundation of China (Grant No.~12274142), and  Introduced Innovative Team Project of Guangdong Pearl River Talents Program (Grant No. 2021ZT09Z109).
A.F.K.~acknowledges support from the Swedish Research Council (grant 
number 
2019-03696), the Swedish Foundation for Strategic Research (grants 
numbers 
FFL21-0279 and FUS21-0063), the Horizon Europe programme 
HORIZON-CL4-2022-QUANTUM-01-SGA via the project 101113946 
OpenSuperQPlus100, 
and from the Knut and Alice Wallenberg Foundation through the Wallenberg 
Centre for Quantum Technology (WACQT). F.N. is supported in part by the Japan Science and Technology Agency (JST) [via the CREST Quantum Frontiers program Grant No. JPMJCR24I2, the Quantum Leap Flagship Program (Q-LEAP), the Moonshot R\&D Grant Number JPMJMS256E, and the ASPIRE program (Grant Number JPMJAP2513)].

%%%%%%%%%%%%%%%%%%%%%%%%%%%%%%%%%%%%%%%%
\vspace{.3cm}
\noindent{\large \textbf{Author contributions} 
\\
\noindent
X.W. and T.L conceived the original idea. J.Q.L. did the 
analytical and numerical analysis under the supervision of X.W. and 
T.L. Z.H.W., A.F.K., L.D. and F.N. provided very useful 
insights and guidance. All authors 
contributed to and 
approved the final version of
the manuscript.

\vspace{.3cm}
\noindent{\large \textbf{Competing interests} 
\\
\noindent
The authors declare no competing interests.

\vspace{.3cm}
\noindent{\large \textbf{Additional information} 
\\
\noindent{\textbf{Supplementary information} 
The online version contains supplementary material 
available at xxx

\end{document}